\documentclass[aps,prd,amsmath,floats,floatfix,twocolumn, nofootinbib,superscriptaddress,showpacs]{revtex4-1}
\usepackage{eurosym}
\usepackage{amsmath}
\usepackage{amsfonts}
\usepackage{array}
\usepackage{amsthm}
\usepackage{bm}
\usepackage{graphicx}
\usepackage{color}
\usepackage{latexsym}
\usepackage[colorlinks,linkcolor=blue,citecolor=blue,urlcolor=black]{hyperref}

\newcommand{\be}{\begin{eqnarray}}
\newcommand{\ee}{\end{eqnarray}}

\newcommand{\bwide}{\begin{widetext}}
	\newcommand{\ewide}{\end{widetext}}
\newcommand{\dee}{\,\textrm{d}}

\newcommand{\bead}{\begin{eqnarray}\begin{aligned}}
\newcommand{\eead}{\end{aligned}\end{eqnarray}}

\begin{document}

	\title{Shadow, quasi-normal modes and quasi-periodic oscillations of rotating Kaluza-Klein black holes}

	\author{M. Ghasemi-Nodehi}
	\email{mghasemin@ipm.ir}
	\affiliation{School of Astronomy, Institute for Research in Fundamental Sciences (IPM), P. O. Box 19395-5531, Tehran, Iran}
	
	\author{Mustapha Azreg-A\"{i}nou}
	\email{azreg@baskent.edu.tr}
	\affiliation{Ba\c{s}kent University, Engineering Faculty, Ba\u{g}l{\i}ca Campus, 06790-Ankara, Turkey}
	
	\author{Kimet Jusufi}
	\email{kimet.jusufi@unite.edu.mk}
	\affiliation{Physics Department, State University of Tetovo, Ilinden Street nn, 1200,
		Tetovo, North Macedonia}
	\affiliation{Institute of Physics, Faculty of Natural Sciences and Mathematics, Ss. Cyril
		and Methodius University, Arhimedova 3, 1000 Skopje, North Macedonia}
	
	\author{Mubasher Jamil}
	\email{mjamil@zjut.edu.cn (corresponding author)}
	\affiliation{Institute for Theoretical Physics and Cosmology, Zheijiang University of Technology, Hangzhou 310023, China}
	\affiliation{Department of Mathematics, School of Natural Sciences (SNS), National
		University of Sciences and Technology (NUST), H-12, Islamabad 44000, Pakistan}
	\affiliation{Canadian Quantum Research Center 204-3002 32 Ave Vernon, BC V1T 2L7, Canada}
	
	\date{\today}

\begin{abstract}
	In this paper we study the shadow of rotating KK black holes and the connection between the shadow radius and the real part of quasi-normal modes (QNMs) in the eiokonal limit. In addition we have explored the  quasi-periodic oscillations (QPOs) in the rotating KK black hole.
\end{abstract}


	\maketitle

\section{Introduction}

Black holes are the standard testbeds for verifying the effects and predictions of strong gravity.
There has been a long standing interest among researchers to find new black hole solutions in alternative and higher dimensional gravitational theories. With the recent experimental evidence of black holes, such as the detection of the gravitational waves \cite{AbbottBH} and the images of the M87 black hole shadow a new window in our understanding of the Universe has been open \cite{EHT1,EHT2,EHT3,EHT4,EHT5,EHT6}. In particular  we can test  many of alternative theories of gravity in the near future based on such astronomical observations.  During the evolution of binary black holes there is the inspiral stage \cite{Blanchet}, the merger phase  \cite{Pretorius,Campanelli,Baker} and finally  the
ringdown phase which describes a perturbed BH that emits GWs in the form of
quasinormal radiation \cite{BertiCardosoWill}. During this stage the black hole emits gravitational waves which can be studied in terms of the quasinormal modes (QNMs). Among other things, QNMs encodes valuable information about the black hole stability under small perturbations,  perturbation theory of black hole and its stability under small perturbations. The problem of QNMs concerning the stability of different black hole solutions have been investigated in many studies using various numerical methods \cite{Regge,Zerilli,1,2,3,4,5,6,7,8,9,10,11,12,13,14,15,16}.

With the recent EHT image of M87 black hole, the problem of black hole shadow has recently gained a lot of attention. Historically the shadow of Schwarzschild BH was studied by Synge \cite{Synge66} and Luminet \cite{Luminet79} and then subsequently the Kerr BH's shadow was studied by Bardeen \cite{DeWitt73}. In recent years, several studies have been performed concerning shadows of rotating black holes in different gravitational theories \cite{r1,r2,r3,r4,r5,r6,r7,r8,r9,r10,myshadow1,myshadow2,000,01,00,111,22,33,44,55,66,77,88,99,100,1111,222,333,444,Banerjee:2019nnj, Feng:2019zzn,Zhang:2019glo}. It turn out that there exists a connection between the circular null geodesics and the real part of QNMs in the eikonal limit as was firstly argued  by Cardoso et al. \cite{cardoso}, see also \cite{Hod:2017xkz,Konoplya:2017wot,Wei:2019jve}. This connection was extended in the strong deflection limit by Stefanov et al. \cite{Stefanov:2010xz}. In Ref. \cite{Jusufi:2019ltj}, one of the authors of this paper argued that one can relate the real part of the QNMs with the shadow radius in the eikonal limit  \cite{Jusufi:2019ltj} (see, also \cite{Liu:2020ola}). Very recently, this connection has been extended for rotating and asymptotically flat black holes as well \cite{Jusufi:2020dhz}.

 The simplest version of the five dimensional Kaluza-Klein theory is to study general relativity in five dimensions and subsequently dimensionally reducing the action in four dimensions. Horne and Horowitz discovered a new black hole solution in the four dimensional Einstein-Maxwell-dilaton theory which is a low-energy effective version of string theory. For particular values of the coupling parameters $\gamma=0$ and $\gamma=\sqrt{3}$, the corresponding solutions are Kerr-Newman and the Kaluza-Klein rotating black holes, respectively \cite{horne}. The solution technique involved a dimensional reduction of the boosted five dimensional Kaluza-Klein solution to four dimensions. Here mass, charge and angular momentum of the black hole are expressed as a function of the boost velocity. The shadow of rotating Kaluza-Klein dilaton black hole was investigated in \cite{eiroa} and the authors found that the dilaton field causes the shadow to be slightly bigger and a reduced deformity as compared with the Kerr black hole.
Wang explored another rotating Kaluza-Klein black hole solution with squashed 3-sphere $S^3$ horizons in five dimensions \cite{twang}. The underlying geometrical structure asymptotically is a twisted $S^1$ fiber bundle over a four dimensional Minkowski spacetime. The derivation involved a squashing transformation containing the inner and outer horizons along with $r_\infty$ characterizing the size of the $S^1$ fiber at infinity, and a five dimensional Kerr black hole with two equal angular momenta. It was shown that the resulting five dimensional squashed Kaluza-Klein black hole is a vacuum solution of the Einstein field equations. The author showed further that the spacetime is geodesically complete and free from naked singularities. Later, Long et al, investigated the properties of shadows of the rotating squashed Kaluza-Klein black hole and compared them with the typical rotating black hole \cite{long}. They demonstrated that for certain choices of the parameters, the black hole has no shadow due to the role of specific angular momentum of photons from the fifth dimension. They also observed that the size of black hole shadow radius increases or decreases monotonically with respect to parameters of the squashed geometry and the fifth dimension.
Larsen derived the most general black hole solutions to the Einstein theory in five dimensions and later dimensionally reducing the solution to four dimensions \cite{fin}. The black hole solution contained four free parameters namely, mass, electric and magnetic charge and the spin. Under the special values of parameters, the four dimensional black hole solution reduces to the Kerr-Newman spacetime but never reduces to Kerr solution. In the framework of M-theory, the electric Kaluza-Klein charge is identified as the D0 brane charge while the magnetic Kaluza-Klein charge becomes the charge of the D6 brane. Larsen suggested that consequently the black hole can be interpreted at the weak coupling of the D0 and D6 branes.

Recently, some of the present authors studied the observational constraints on the parameters of the four dimensional Kaluza-Klein black hole via X-ray reflection spectroscopy technique \cite{xray}. From this point of view, the observational constraints on the parameters of the four dimensional Kaluza-Klein black hole could also be determined via theoretical calculations of the quasi-periodic oscillations (QPOs) and curve fitting to existing experimental data for rotating microquasars. The appearance of two peaks at 300 Hz and 450 Hz in the X-ray power density spectra of Galactic microquasars, representing possible occurrence of a lower QPO and of an upper QPO in a ratio of 3 to 2, has stimulated a lot of theoretical works to explain the value of the 3/2-ratio. Some theoretical models, including parametric resonance, forced resonance and Keplerian resonance have been proposed. In this paper we rely on the parametric resonance model for our investigations. Thus, in this work we will be using two different tools, shadow and QPOs, to determine the bounds (limits) of the parameters of the four dimensional Kaluza-Klein black hole. In the QPOs method we describe a microquasar as a four dimensional Kaluza-Klein black hole and we have selected three microquasars the astrophysical data of which are the most accurate.

This paper is organized as follows. In Sec.~\ref{secII}, we review the rotating Kaluza-Klein black hole metric. In Sec.~\ref{sha}, we determine the shadow of the Kaluza-Klein rotating black hole. In Sec.~\ref{res}, we discuss the results of the Kaluza-Klein shadow. In Sec.~\ref{seccon}, we study the connection between the shadow radius and the QNMs in the eiokonal regime. Section~\ref{secqpos} is devoted to the investigation of the Kaluza-Klein black hole QPOs and their relations to the observed data for three rotating microquasars. Finally, in Sec.~\ref{summary} we comment on our results.

\section{Rotating KK black hole\label{secII}}

KK theories are of tremendous interest in string theory community because of their roles as low-energy approximations to string theory. As mentioned earlier, Larsen derived the most general black hole solutions to the Einstein theory in five dimensions and later dimensionally reducing the solution to four dimensions. The solution generating technique of the following KK black hole has been discussed in \cite{fin} and the resulting spacetime metric is
\bwide
\bead\label{solution}\label{5d}
	{\rm d}s^2=\frac{H_2}{H_1}({\rm d}\psi +A)^2-\frac{H_3}{H_2} ({\rm d}t+B)^2+H_1 \Big(  \frac{{\rm d}r^2}{\Delta}+{\rm d}\theta^2+\frac{\Delta}{H_3}\sin^2\theta \dee\phi^2 \Big),
\eead
where the one-forms are given by
\bead
	A=& - \frac{1}{H_2}\Big[  2Q(r+\frac{p-2m}{2})+\sqrt{\frac{q^3(p^2-4m^2)}{4m^2 (p+q)}}a\cos\theta  \Big]\dee t
	-\frac{1}{H_2}\Big[  2p(H_2+a^2\sin^2\theta)\cos\theta\\
	&+\sqrt{\frac{p(q^2-4m^2)}{4m^2 (p+q)^3}}[(p+q)(pr-m(p-2m))+q(p^2-4m^2)]a\sin^2\theta  \Big]\dee\phi,\\
	\label{B3}B=&\frac{(pq+4m^2)r-m(p-2m)(q-2m)}{2m(p+q)H_3}\sqrt{pq}a\sin^2\theta \dee\phi,
\eead
and
\bead\label{eq-5dmetricfuncs}
	H_1 &= r^2 +a^2\cos^2\theta +r(p-2m)+\frac{p(p-2m)(q-2m)}{2(p+q)}-\frac{p}{2m(p+q)}\sqrt{(q^2-4m^2)(p^2-4m^2)}~a\cos\theta,\\
	H_2 &= r^2 +a^2\cos^2\theta +r(q-2m)+\frac{q(p-2m)(q-2m)}{2(p+q)}+\frac{q}{2m(p+q)}\sqrt{(q^2-4m^2)(p^2-4m^2)}~a\cos\theta,\\
	H_3 &= r^2+a^2\cos^2\theta-2mr,\\
	\Delta &= r^2+a^2-2mr.
\eead
\ewide
The solution admits four free parameters, viz. $m,a,p,q$, which are related to the physical mass $M$, the angular momentum $J$ and the electric ($Q$) and magnetic charge ($P$) respectively. The relations are given as \cite{fin}:
\bead\label{eq-rels}
M = \frac{p+q}{4}, ~
J = \frac{\sqrt{pq}(pq+4m^2)}{4m(p+q)}a, \\
Q^2 = \frac{q(q^2-4m^2)}{4(p+q)}, ~
P^2 = \frac{p(p^2-4m^2)}{4(p+q)}.
\eead
It needs to be stressed that the above metric does not reduce to Kerr BH by setting $p=0=q$, which ultimately leads to vanishing or indeterminate forms of physical parameters. To get a meaningful reduction, at least one charge must be nonzero which than lead to a metric similar to the Kerr Newman black hole.

In five-dimensional Kaluza-Klein theories the spacetime is equipped with a five dimensional metric $g_{\alpha\beta}$ independent of the extra spacelike dimension $x^5=\psi$~\cite{KK}: ${\rm d}s^2=g_{\alpha\beta}(x^{a}){\rm d}x^{\alpha}{\rm d}x^{\beta}$, of signature ($+,+,+,-,+$). Here ($\alpha,\,\beta$) run from 1 to 5 and Latin indexes ($a,\,b$) run from 1 to 4.

The compactification of the fifth dimension is performed using the dimensional reduction which converts the five dimensional black string into a four-dimensional BH with the metric \cite{fin,MMK}:
\bwide
\begin{equation}\label{eq-4dmetric}
\dee s^2 = -\frac{H_3}{\rho^2}\dee t^2 -2\frac{H_4}{\rho^2}\dee t \dee \phi + \frac{\rho^2}{\Delta}\dee r^2 + \rho^2\dee \theta^2 + \left ( \frac{-H_4^2 + \rho^4 \Delta \sin^2{\theta}}{\rho^2H_3} \right )\dee \phi^2 ,
\end{equation}
where $\rho^2 = \sqrt{H_1 H_2}$ and
\begin{eqnarray}
\frac{H_1}{M^2}&=& \frac{8(b-2)(c-2)b}{(b+c)^3}+\frac{4(b-2)x}{b+c}+x^2-\frac{2b\sqrt{(b^2-4)(c^2-4)}~\alpha\cos\theta}{(b+c)^2}+\alpha^2\cos^2\theta,\nonumber\\
\frac{H_2}{M^2}&=& \frac{8(b-2)(c-2)c}{(b+c)^3}+\frac{4(c-2)x}{b+c}+x^2+\frac{2c\sqrt{(b^2-4)(c^2-4)}~\alpha\cos\theta}{(b+c)^2}+\alpha^2\cos^2\theta,\nonumber\\
\label{eq-4dmetricfuncs}\frac{H_3}{M^2}&=& x^2+\alpha^2\cos^2\theta-\frac{8x}{b+c},\\
\frac{H_4}{M^3}&=&\frac{2\sqrt{bc}[(bc+4)(b+c)x-4(b-2)(c-2)]\alpha\sin^2\theta}{(b+c)^3},\nonumber\\
\frac{\Delta}{M^2}&=&x^2+\alpha^2-\frac{8x}{b+c}.\nonumber
\end{eqnarray}
\ewide
Here the dimensionless parameters were defined as $\alpha \equiv a/M, b \equiv p/m, c \equiv q/m$, and $x \equiv r/M$. Furthermore $m$ and $M$ are also related by \be\label{eq-mM}
m = 4M/(b+c).
\ee
Note that the the spin parameter $\alpha$ is not always the same as the dimensionless spin parameter of the Kerr metric. Only when the electric and magnetic charges are zero, and the Kaluza-Klein metric reduces to the Kerr metric, does $\alpha$ equal the $a_*$ parameter of the Kerr solution.  The spacetime admits two horizons
\be
r_{\pm} = m \pm \sqrt{m^2-a^2},
\ee
or, in terms of the dimensionless quantities,
\be\label{eq-xhorizon}
x_{\pm} = \frac{4 \pm \sqrt{16-\alpha^2(b+c)^2}}{b+c},
\ee
and the determinant is equal to $\rho^2\sin^2{\theta}$. The Kerr solution is recovered when $b=c=2$.  If we now use the definitions for $Q^2$ and $P^2$ in Eq.~\ref{eq-rels}, we have the conditions $q\geq2m,\; p\geq2m$, or $b\geq2,\; c\geq2$. Using these and Eq.~\ref{eq-xhorizon} we arrive at a bound on $\alpha$:
\be\label{eq-spinbound}
	\alpha^2 < 1, \quad \text{or} \quad -1 < \alpha < 1.
\ee
Here $\alpha>0 \;(<0)$ can be interpreted as a counter-rotating (co-rotating) black hole relative to a stationary observer. For upper bounds on $b, c$, we first note that we have set $b=c$. Then, using Eq.~\ref{eq-xhorizon} and Eq.~\ref{eq-spinbound} gives us
\be
	b^2 < \frac{4}{\alpha^2}\, .
\ee
Thus,
\be\label{eq-brange}
	2 \leq b < \frac{2}{|\alpha|}\, .
\ee

\subsection{Shape of the ergoregion}
	
	{We can use KK black hole \eqref{eq-4dmetric} to investigate the horizons and shape of the ergoregion of the spacetime geometry. To find the corresponding horizons of the KK black hole we need to solve the following equation
	\begin{equation}
	\frac{\Delta}{M^2}=x^2+\alpha^2-\frac{8x}{b+c}=0.
	\end{equation}
	On the other hand, the so-called static limit or ergo-surface, inner and outer, is obtained by solving
	\begin{equation}
	\frac{H_3}{M^2}= x^2+\alpha^2\cos^2\theta-\frac{8x}{b+c}=0.
	\end{equation}
	
	By varying the angular momentum parameter $a$, in Fig. \ref{FigE} we depict the effect of the parameters $b$ and $c$ on the the surface horizon and ergoregion, respectively.  We see that there is a domain of parameters and a critical value of $a$ such that the horizons disappear.
	 \begin{figure*}[!htb]
		\includegraphics[width=7.5cm]{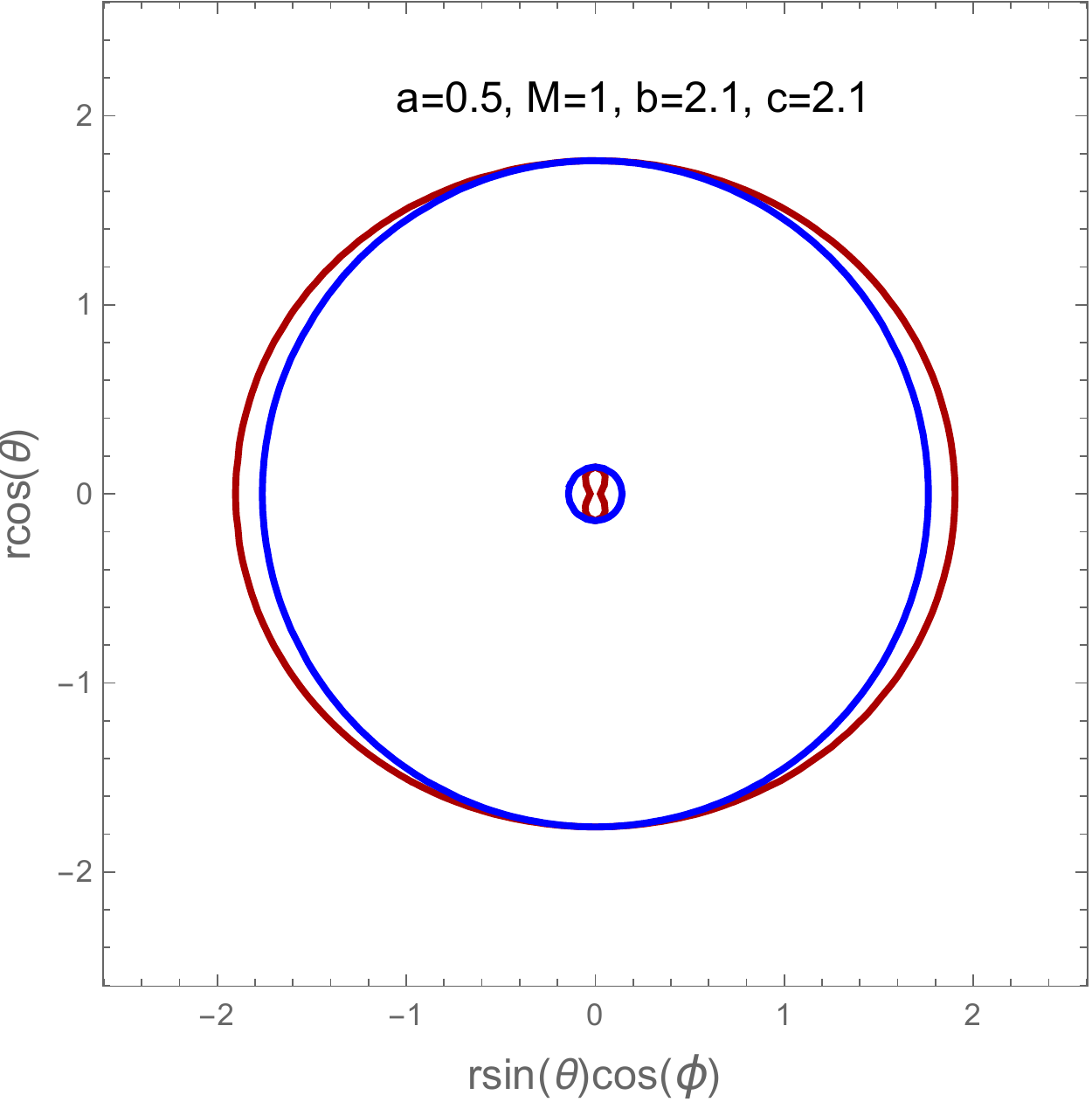}
		\includegraphics[width=7.5cm]{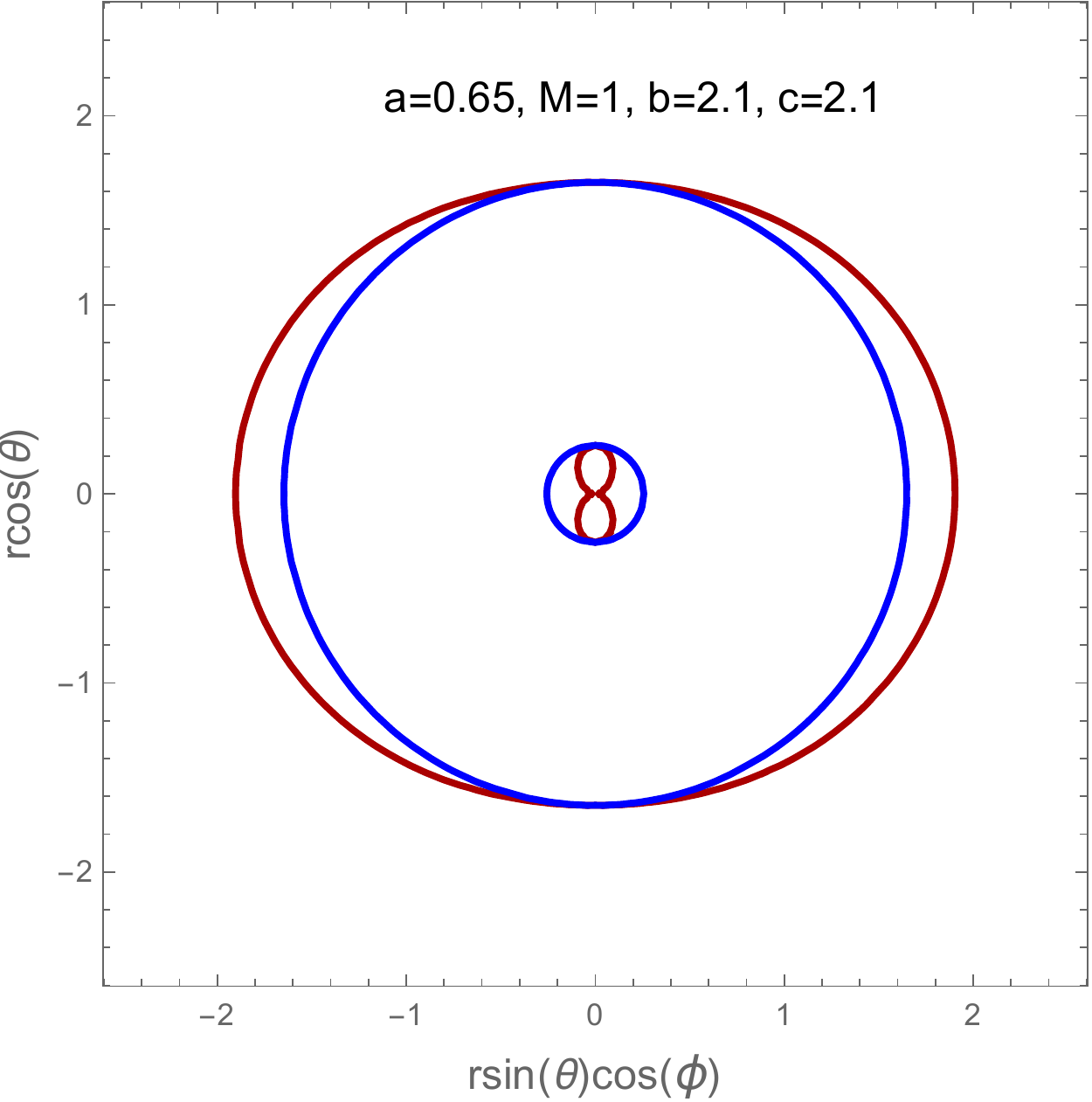}
		\includegraphics[width=7.5cm]{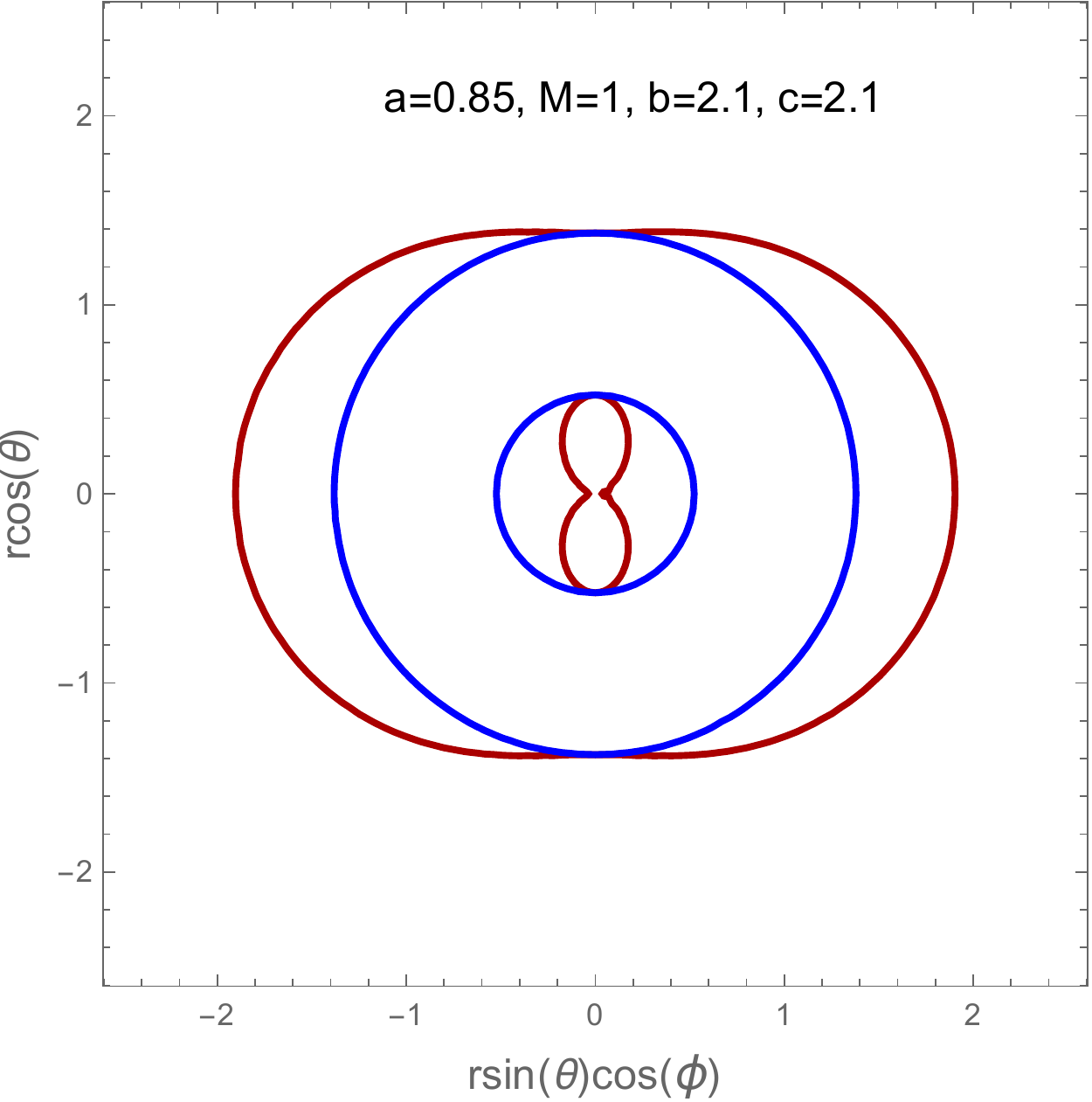}
		\includegraphics[width=7.5cm]{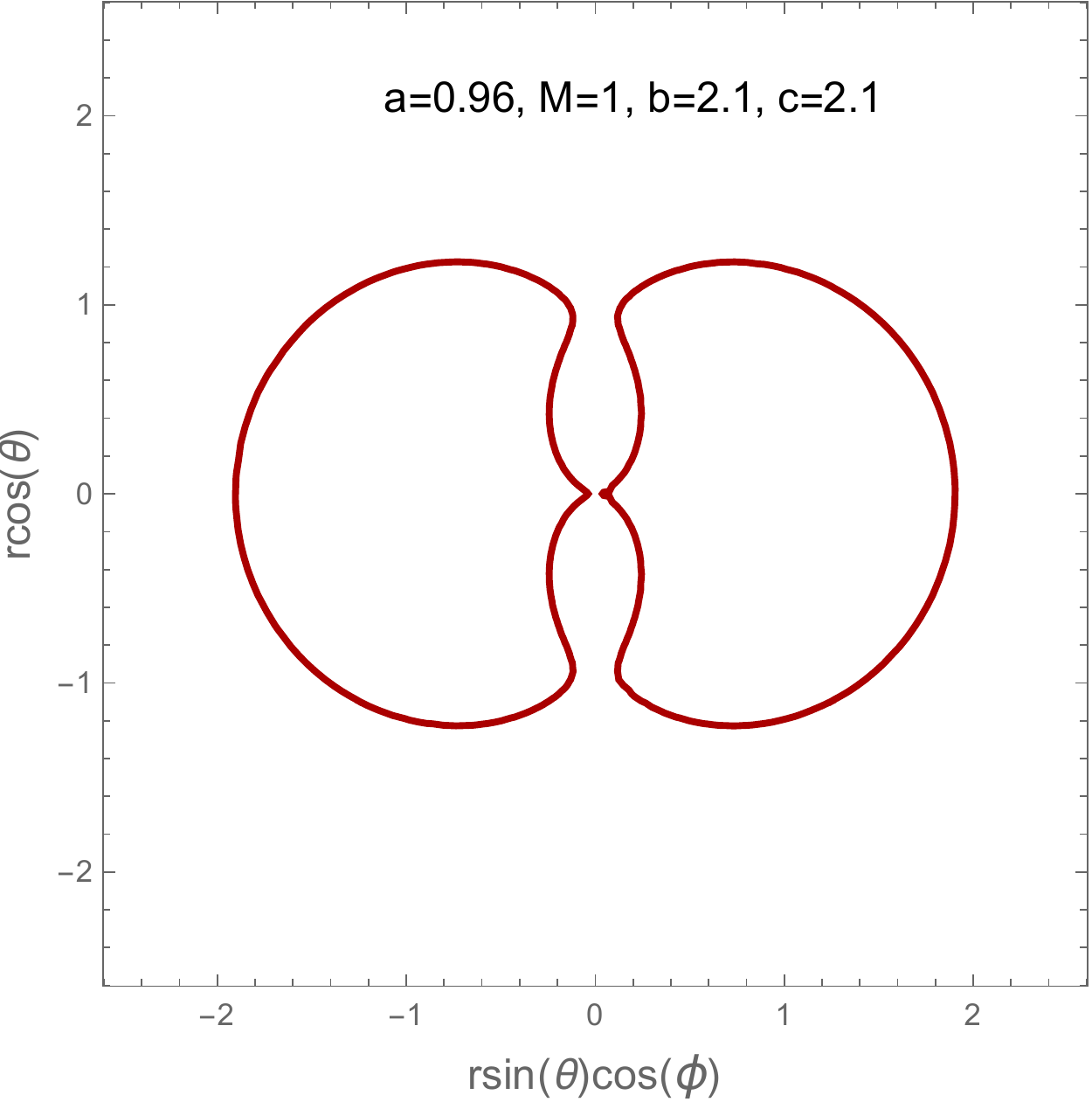}
		\caption{Surface horizon (blue color) and ergoregion (red color) of the KK black hole for different values of $a$. In the domain of parameters used in these plots, we find that the horizon disappears at some critical value $a_c=0.953$.} \label{FigE}
	\end{figure*}
	
\subsection{Embedding Diagram\label{secem}}
	In this section, we investigate the geometry of the KK spacetime, by embedding it into a higher-dimensional Euclidean space. To this purpose, let us consider
		the  equatorial plane $\theta=\pi/2$ at a fixed moment  $t=$ Constant, for which the metric can be written as
	\begin{equation}
	\dee s^2=\frac{\dee r^2}{1-\frac{b(r)}{r}}+\mathcal{R}^2\dee \phi^2, \label{emb}
	\end{equation}
	with
	\begin{eqnarray}
	b(r)&=& r \left(1-\frac{\Delta}{H_1 H_2}\right),\\
	\mathcal{R}(r)&=&\left ( \frac{-H_4^2 + \rho^4 \Delta }{\rho^2H_3} \right )^{1/2}. ~~~~
	\end{eqnarray}
	{Let us embed this reduced  BH metric into three-dimensional Euclidean space  in the cylindrical coordinates,
		\begin{eqnarray}\notag
		\dee s^2&=&dz^2+\dee \mathcal{R}^2+\mathcal{R}^2\dee \phi^2,
		\end{eqnarray}
		which can be further written as
		\begin{eqnarray}\notag
		\dee s^2&=&\left[ \left(\frac{\dee \mathcal{R}}{\dee r}\right)^2+\left(\frac{\dee z}{\dee r}\right)^2  \right]\dee r^2+\mathcal{R}^2\dee \phi^2.
		\end{eqnarray}
		From these equations we obtain
	\begin{equation}
	\frac{\dee z}{\dee r}=\pm \sqrt{\frac{r}{r-b(r)}-\left(\frac{\dee \mathcal{R}}{\dee r}\right)^2},
	\end{equation}
	 In Fig. \ref{FigEM} we show the corresponding KK spacetime embedded in a three-dimensional Euclidean for given value of parameters.
	
	\begin{figure}[!htb]
		\includegraphics[width=7.5cm]{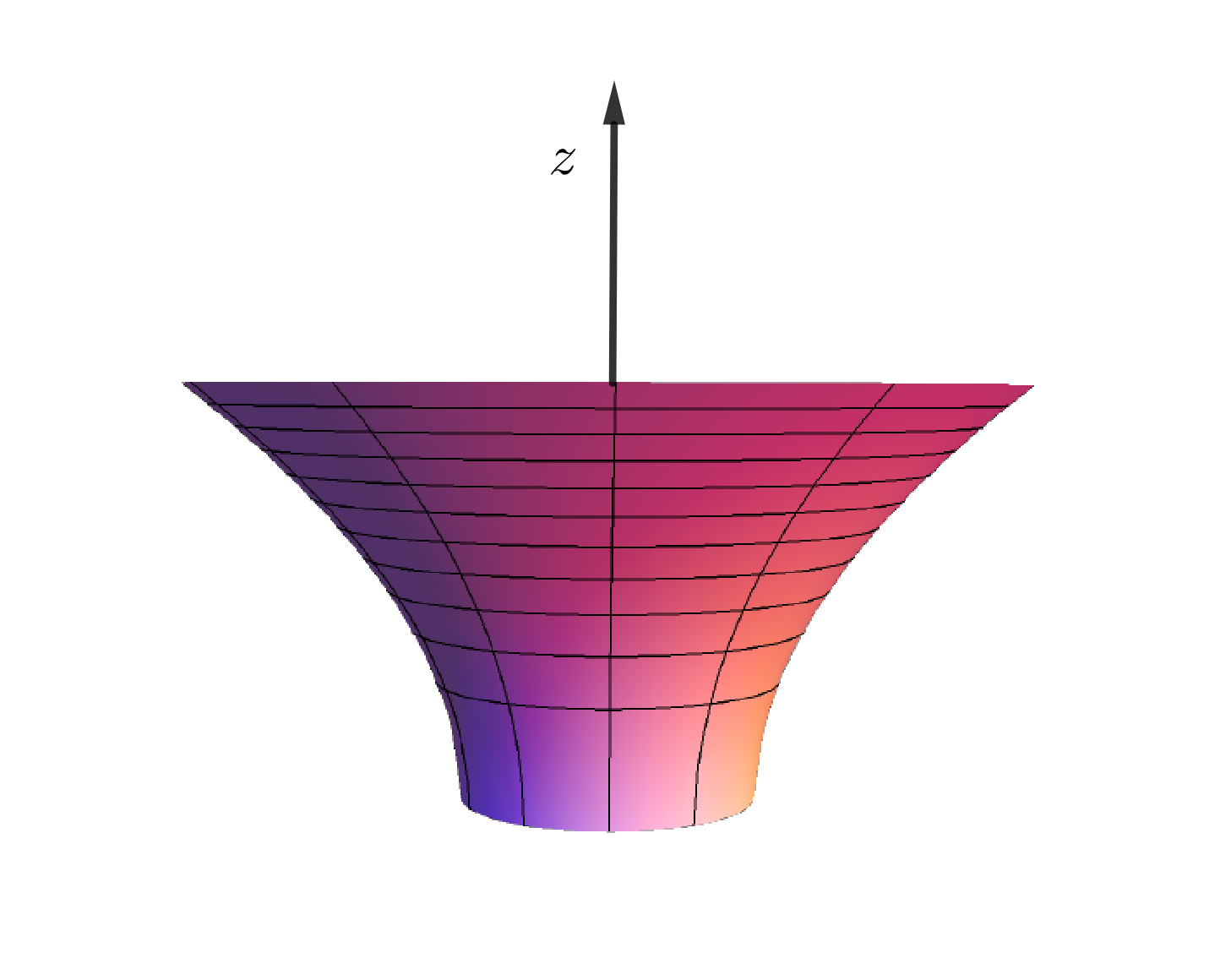}
		\caption{The KK spacetime embedded
			in a three-dimensional Euclidean space. We have set $a=0.5$, $b=2.1$, $c=2$ and $M=1$. }\label{FigEM}
\end{figure}

\section{Shadow boundary \label{sha}}

The photon capture sphere (or unstable photon orbits) of a compact object lies between photon captured orbits and photons scattered orbits, namely the orbits of those photons that are fired from infinity and then are, respectively, captured by the object or scattered back to infinity.  If the compact object is surrounded by an optically thin emitting medium, it turns out that a significant fraction of the photons emitted by the medium can orbit around the compact object and close to the photon capture sphere several times, which generates a brighter boundary and a central dark region on the image plane of the distant observer. Such a dark area is commonly called the black hole shadow and the bright boundary of the black hole shadow corresponds to the apparent image of the photon capture sphere.

Solving geodesic equation is needed to study shadow of compact object. Here, the solution to geodesic equation of rotating KK metric is done numerically using a ray-tracing code. We use simulation that has been done in~\cite{myshadow1} and~\cite{myshadow2}. The solution is based on a class of Runge-Kutta-Nystrom method with adaptive step sizes with error control~\cite{Lund}.
The geometry of system contains two cartesian coordinates as figure 1 of Ref.~\cite{geo}.
One is located at observer plane and another is as coordinate of compact object with distance $D$ and viewing angle $i$. We fire photon from observer plane and photon 3-momentum is perpendicular to observer plane. The photon trajectory integrated from observer plane to the BH with initial condition. The initial condition is presented in the appendix of paper~\cite{geo}.

 To parametrize shadow boundary first we need shadow center. The center is defined as
\be
x'_{\rm C} &=& \frac{\int \rho(x',y') x' dx' dy'}{\int \rho(x',y') dx' dy'} \,  \nonumber\\
y'_{\rm C} &=& \frac{\int \rho(x',y') y' dx' dy'}{\int \rho(x',y') dx' dy'} \, ,
\ee
where inside shadow is presented by $\rho(x',y') = 1$  and outside shadow is $\rho(x',y') = 0$  Symmetry axis of shadow is considered as $x'$ axis. The shorter segment on $x'$ axis is starting point, $\phi =0 $, and $R(\phi)$ is defined as distance between each point of boundary and the center of the shadow.

By the recent shadow observation of M87* of EHT collaborations, the boundary of the black hole shadow does not appreciably deviate from circularity. This permits us to constrain the spacetime metric around black hole from the circularity of its black hole shadow. To calculate circularity of shadow boundary one need to define average radius as
\be
\bar R^2 \equiv \frac{1}{2\pi} \int_0^{2\pi} \! R^2(\phi) \, \mathrm{d}\phi.
\ee

Following~\cite{EHT1} the difference between the RMS distance and the average radius of the shadow can define the deviation from circularity of shadow boundary as follows:
\be
\label{circu}
\Delta C \equiv \frac{1}{ \bar R} \,\,\sqrt{\frac{1}{2\pi} \int_0^{2\pi} \! (R(\phi) - \bar R)^2 \, \mathrm{d}\phi}\,.
\ee

This parametrization can be used to compare theoretical shadow boundaries with observational case. The reported EHT group results for M87* is $\Delta C \lesssim 10\%$~\cite{EHT1}.

Furthermore, The angular size of shadow can be derived from
\be
\label{sizeeq}
\frac{D\delta}{M} \simeq 11.0 \pm 1.5.
\ee
where, from EHT collaboration report~\cite{EHT6} for M87* we consider the distance of M87* as $D =16.8^{+0.8}_{-0.7}$\,Mpc, mass of M87* as $M = (6.5 \pm 0.2|_{\rm stat} \pm
0.7|_{\rm sys}) \times 10^9 M_{\odot}$ and $\delta = (42 \pm 3)\, \mu$arcsec. For simplicity,
we consider $D=16.8\pm 0.75$~Mpc in our calculation.

\section{Results and Discussion}\label{res}

We simulate the shadow boundary of spacetime with Kaluza-Klein mteric Eq.~\ref{eq-4dmetric}. Figure~\ref{shape0.9} shows the shadow boundary for different set of parameters $b$ and $c$. The $\alpha$ is 0.9 and viewing angle is chosen to be $17^{\circ}$. We choose viewing angle as  $17^{\circ}$ because this value is reported by~\cite{jet} that is estimated angle between the approaching jet and line of sight. By increasing (decreasing) parameters $b/c$ for the case $b=c$ the shadow size decreases (increases). For the case $b=2$ and $c$ as free parameter, by increasing (decreasing) the value $c$ shadow sizes decreases (increases). For the third case, $c=2$ and b as free parameter, the shadow size decreases (increases) by increasing (decreasing). One sample of shadow size vs parameter $c$ is shown in figure~\ref{size}.
Figure~\ref{shape-0.9} is similar to figure \ref{shape0.9} but the spin parameter is $-0.9$. The shadow size variations is similar to positive spin value.

\begin{figure}
\includegraphics[width=8.0cm]{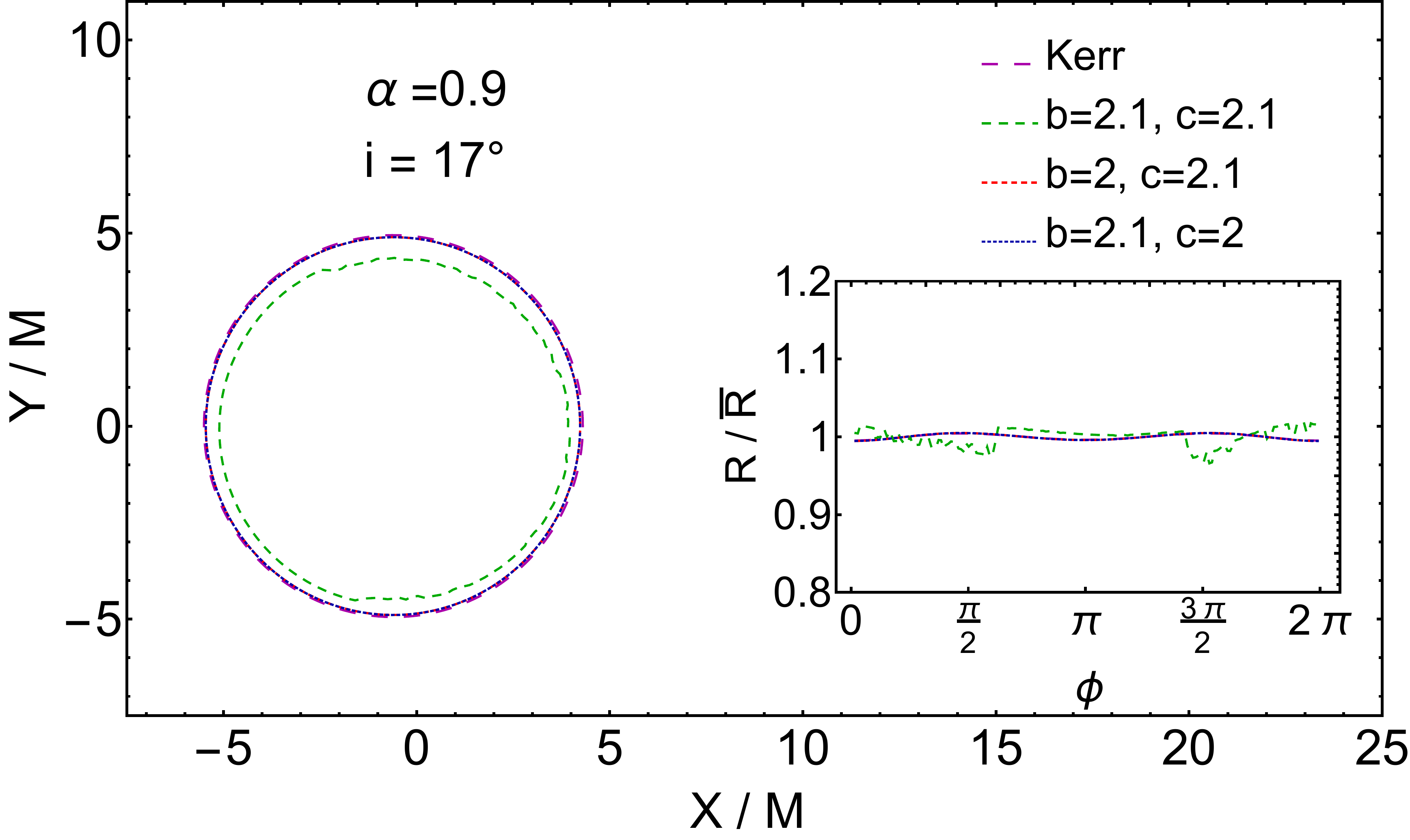}
\caption{This figure shows the shadow boundary for different set of parameters $b$ and $c$. The spin parameter is 0.9 and the viewing angle is $17^{\circ}$. \label{shape0.9}}
\end{figure}

\begin{figure}
\includegraphics[width=8.0cm]{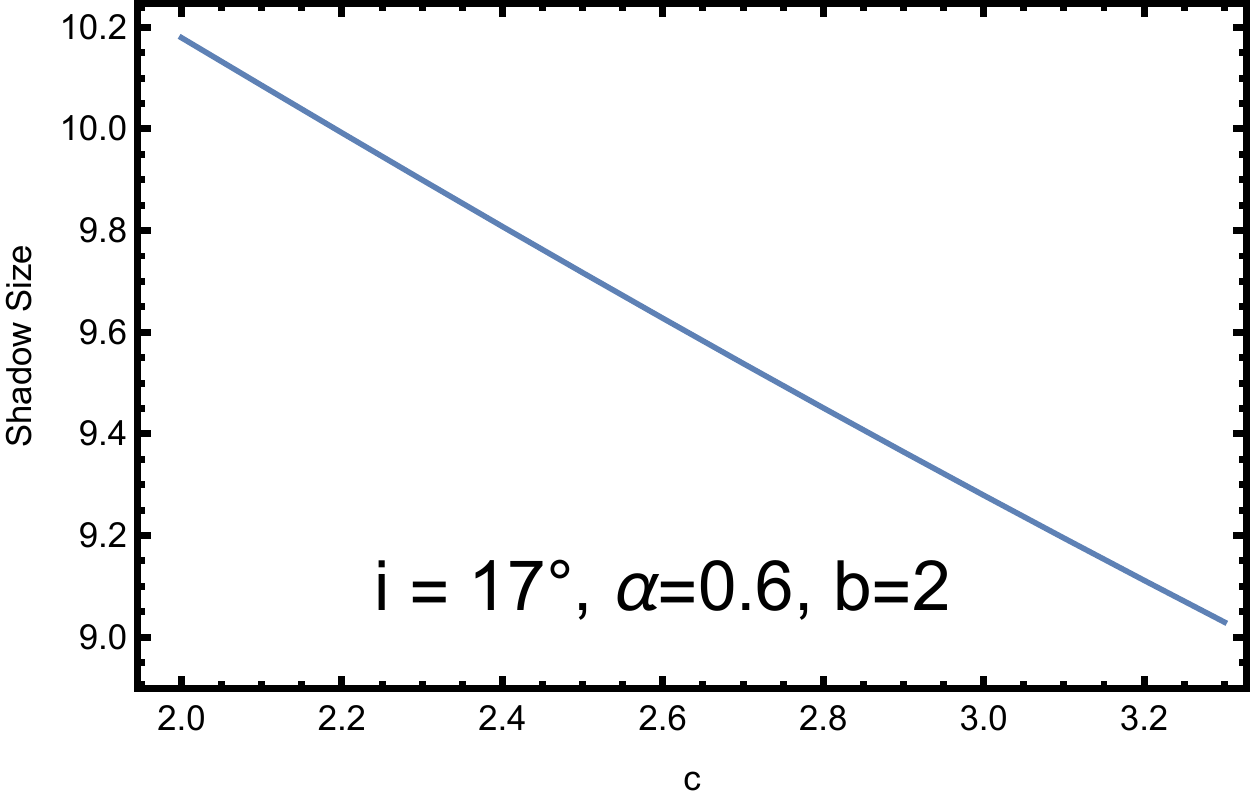}
\caption{Shadow size variations vs parameter $c$ for the case $b=2$. The spin parameter is 0.6 and viewing angle is $17^{\circ}$. \label{size}}
\end{figure}

\begin{figure}
\includegraphics[width=8.0cm]{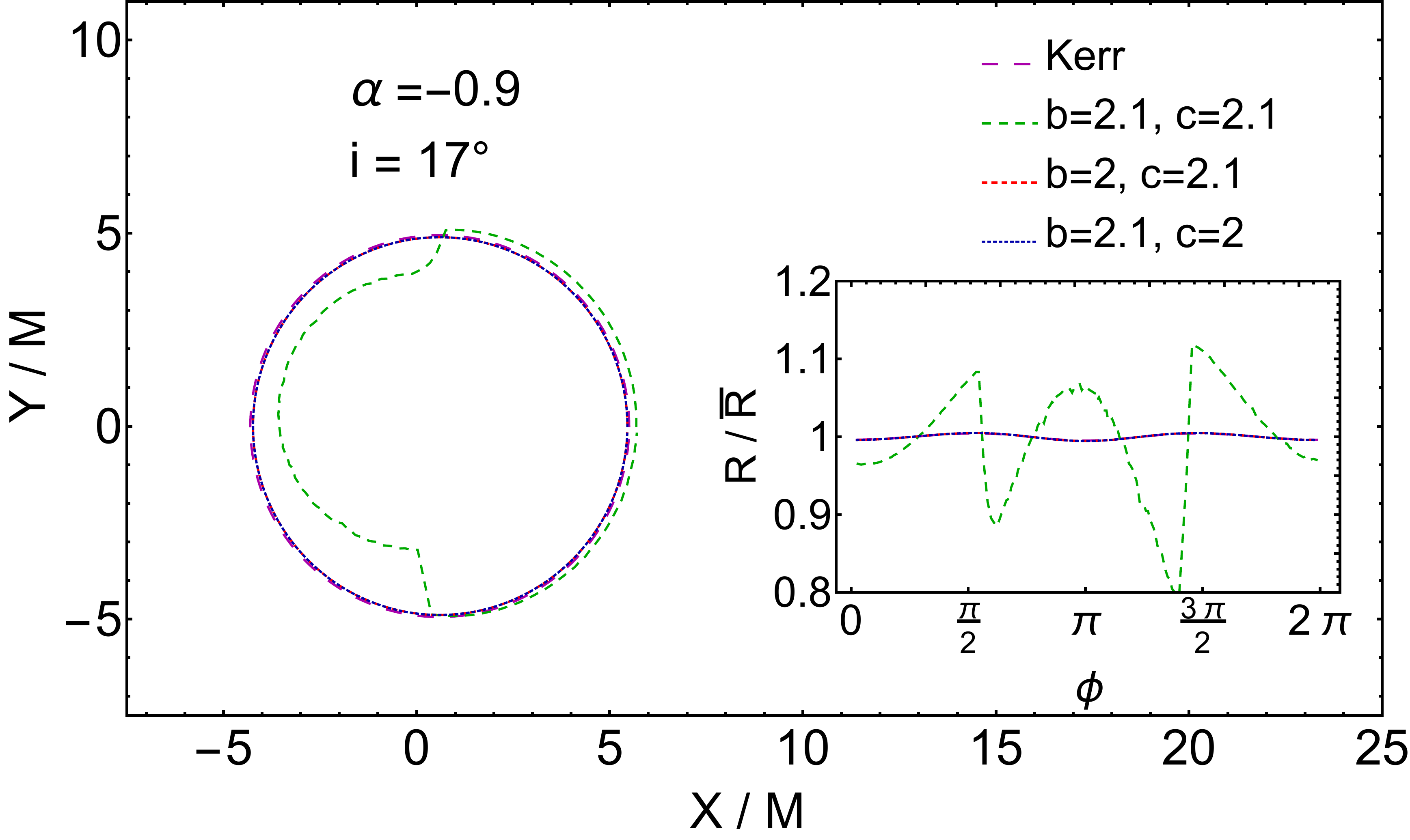}
\caption{This figure shows the shadow boundary for different set of parameters $b$ and $c$. The spin parameter is $-0.9$ and the viewing angle is $17^{\circ}$. \label{shape-0.9}}
\end{figure}

Figures~\ref{bc},~\ref{b2} and~\ref{c2} are drawn to show the combination of $\alpha$ and $b/c$ parameters. Considering rotating KK metric of Eq.~\ref{eq-4dmetric} and observational parameters $\Delta C$ and shadow size of Eq.~\ref{sizeeq}, we can put constraint on parameters $b/c$. The colored region represent deviation from circularity which we choose less than $10\%$ here. The hatched region shows allowed size of shadow according to Eq.~\ref{sizeeq}. The grey region is excluded by limitation explained in Eq.~\ref{eq-brange}. The green dashed line is the boundary of forbidden region from Eq.~\ref{eq-brange}. The white region in Fig.~\ref{bc} is excluded by circularity condition that means the circularity of this region is more than $10\%$. From circularity condition of M87*, $\Delta C \lesssim 10\%$, the $b/c$ parameters greater than 8.9 is not allowed for the case $b=c$. From shadow size constraints $b/c$ in the case $b=c$ can be maximum 2.4 for positive $\alpha$  and 3.7 for negative $\alpha$ in figure~\ref{bc}.

The figure~\ref{b2} is for the case $b=2$ and $c$ is as free parameters. The circularity is below $10\%$ for all allowed region and from shadow size constraints the maximum value for $c$ is 3.1.

The figure~\ref{c2} is for the case $c=2$ and $b$ as free parameter. The circularity is below $10\%$ for all allowed region and shadow size constraints says parameter $c$ cannot be larger than 3.1 for M87*. In all cases we exclude non-rotating case.

\begin{figure}
\vspace{0.4cm}
\begin{center}
\includegraphics[width=8.0cm]{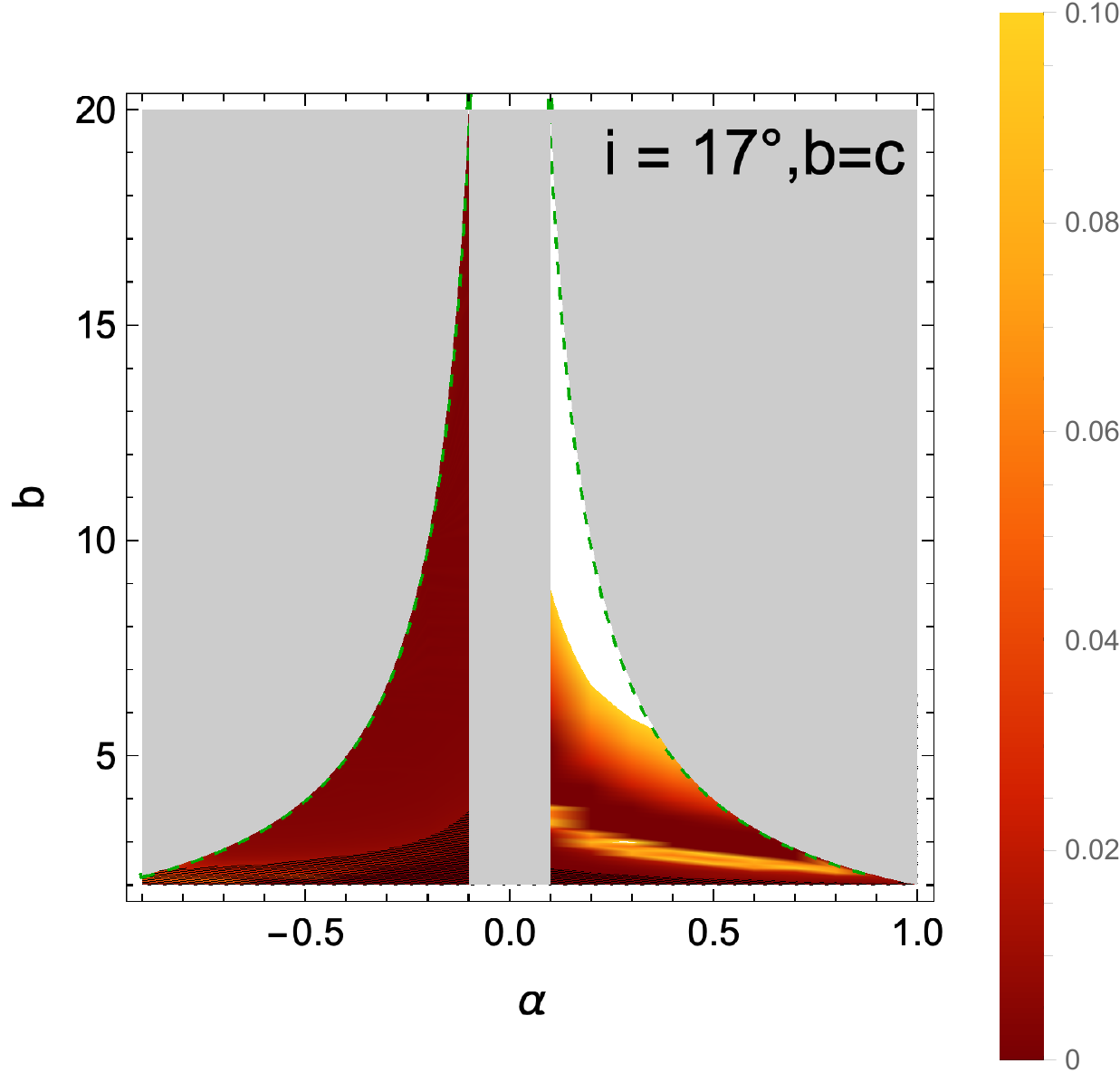}
\end{center}
\vspace{-0.3cm}
\caption{The colored region represent deviation from circularity which we choose less than $10\%$ here. The hatched region shows allowed size of shadow according to Eq.~\ref{sizeeq}. The grey region is excluded by limitation explained in Eq.~\ref{eq-brange}. The green dashed line is the boundary of forbidden region from Eq.~\ref{eq-brange}. The white region is excluded by circularity condition that means the circularity of this region is more than $10\%$. We exclude non-rotating case\label{bc}}
\end{figure}

\begin{figure}
\vspace{0.4cm}
\begin{center}
\includegraphics[width=8.0cm]{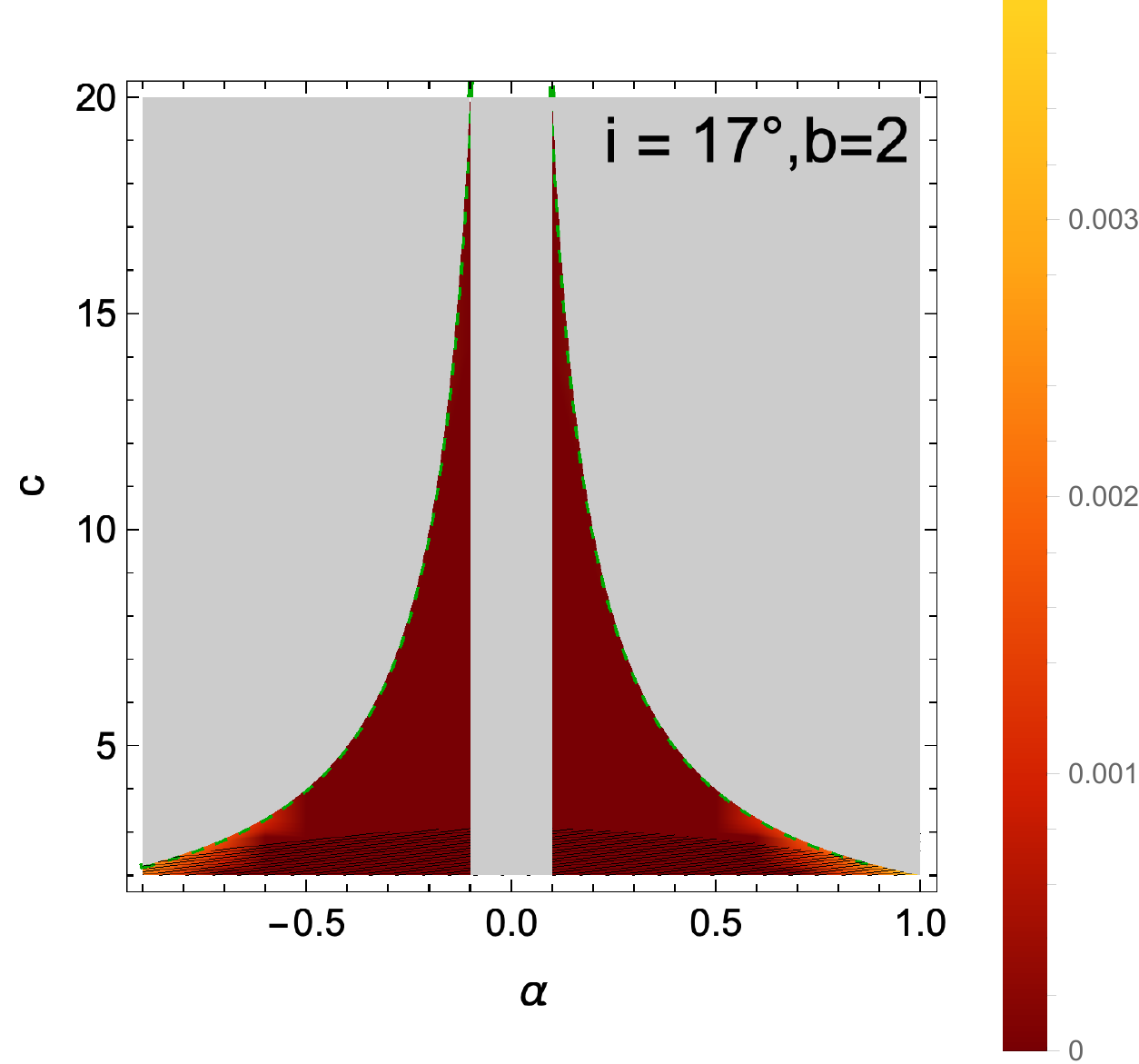}
\end{center}
\vspace{-0.3cm}
\caption{The colored region represent deviation from circularity which we choose less than $10\%$ here. The hatched region shows allowed size of shadow according to Eq.~\ref{sizeeq}. The grey region is excluded by limitation explained in Eq.~\ref{eq-brange}. The green dashed line is the boundary of forbidden region from Eq.~\ref{eq-brange}. The circularity is below $10\%$ here. We exclude non-rotating case.\label{b2}}
\end{figure}

\begin{figure}
\vspace{0.4cm}
\begin{center}
\includegraphics[width=8.0cm]{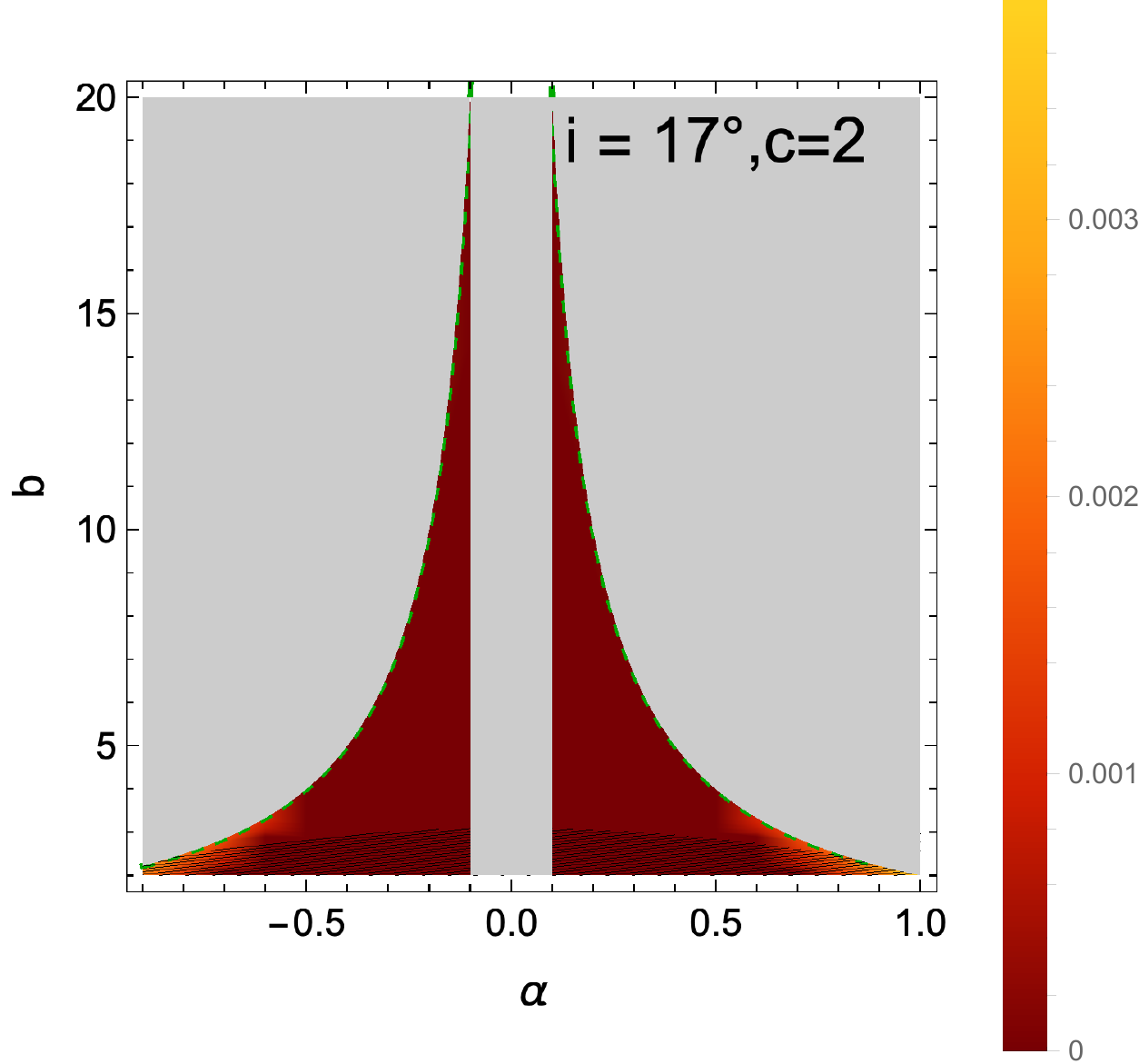}
\end{center}
\vspace{-0.3cm}
\caption{The colored region represent deviation from circularity which we choose less than $10\%$ here. The hatched region shows allowed size of shadow according to Eq.~\ref{sizeeq}. The grey region is excluded by limitation explained in Eq.~\ref{eq-brange}. The green dashed line is the boundary of forbidden region from Eq.~\ref{eq-brange}. The circularity is below $10\%$ here. We exclude non-rotating case.\label{c2}}
\end{figure}

\section{Connection between shadow radius and real part of QNMs}\label{seccon}

The corresponding four dimensional metric in the coordinates $(t,r,\theta,\phi)$ in the in the equatorial planer having $\theta=\pi/2$ yielding
\begin{equation}\label{met}
	{\rm d}\bar{s}^2|_{\theta=\pi/2}=g_{tt}{\rm d}t^2+2 g_{t \phi}{\rm d}t {\rm d}\phi+g_{rr} {\rm d}r^2+
	g_{\phi \phi}{\rm d}\phi^2,
\end{equation}
where
\begin{eqnarray}
g_{tt} &=& -\frac{H_3}{\rho^2},\\
g_{t \phi} &=& -\frac{H_4}{\rho^2},\\
g_{rr} &=& \frac{\rho^2}{\Delta}, \\
g_{\phi \phi} &=& \Big(  \frac{-H_4^2+\rho^4 \Delta }{\rho^2 H_3} \Big).
\end{eqnarray}

Here we have set $\rho^2\equiv \sqrt{H_1 H_2}$ along with the dimensionless parameters ($b,\,c$) where $p\equiv bm$ and $q\equiv cm$, and other dimensionless parameters defined by $\epsilon^2\equiv Q^2/M^2$, $\mu^2\equiv P^2/M^2$, $\alpha\equiv a/M$ and $x\equiv r/M$. From now on we will adopt ($x,\,M,\,\alpha,\,b,\,c$) as free independent parameters in terms of which the relevant quantities take the following form.
\begin{eqnarray}
	\frac{H_1}{M^2}&=& \frac{8(b-2)(c-2)b}{(b+c)^3}+\frac{4(b-2)x}{b+c}+x^2,\\
	\frac{H_2}{M^2}&=& \frac{8(b-2)(c-2)c}{(b+c)^3}+\frac{4(c-2)x}{b+c}+x^2,\\
	\label{r1D}\frac{H_3}{M^2}&=& x^2-\frac{8x}{b+c},\qquad \frac{\Delta}{M^2} = x^2+\alpha^2-\frac{8x}{b+c},\\
	\label{r2D}\frac{H_4}{M^3}&=&\frac{2\sqrt{bc}[(bc+4)(b+c)x-4(b-2)(c-2)]\alpha}{(b+c)^3}
\end{eqnarray}

The appropriate Lagrangian is written as
\begin{equation}
\mathcal{L}=\frac{1}{2}\left(g_{tt}\dot{t}^2+g_{rr} \dot{r}^2+2 g_{t \phi} \dot{t} \dot{\phi}+g_{\phi \phi}\dot{\phi}^2\right).
\end{equation}

The generalized momenta following from this Lagrangian are
\begin{eqnarray}
p_t&=&g_{tt}\dot{t}+g_{t \phi} \dot{\phi}=-E\\
p_{\phi}&=&g_{t\phi}\dot{t}+g_{\phi \phi} \dot{\phi}=J\\
p_r&=&g_{rr}\dot{r}
\end{eqnarray}

Now let us use the Hamiltonian
\begin{equation}
\mathcal{H}=p_t \dot{t}+p_{\phi }\dot{\phi}+p_{r} \dot{r}-\mathcal{L},
\end{equation}
along with the conditions for the existence of circular geodesics  written as follows
\begin{equation}
\label{Vreq}
\mathcal{V}_r=\mathcal{V}^{'}_{r}=0.
\end{equation}
From now on, a prime denotes derivative with respect to radial coordinate $r$ and $\mathcal V$ denotes the effective potential energy of photons.

We therefore obtain the following relations
\begin{equation}\label{Eq30}
g_{\phi \phi}|_{r_0}E^2+2 g_{t \phi}|_{r_0} E J+g_{t t }|_{r_0} J^2=0,
\end{equation}
and
\begin{equation}
g'_{\phi \phi}|_{r_0}E^2+2 g'_{t \phi}|_{r_0} E J+g'_{tt}|_{r_0} J^2=0.
\end{equation}
respectively. Now let us introduce the following quantity
\begin{equation}
R_s=\frac{J}{E}
\end{equation}

\begin{figure*}
\includegraphics[width=7.4cm]{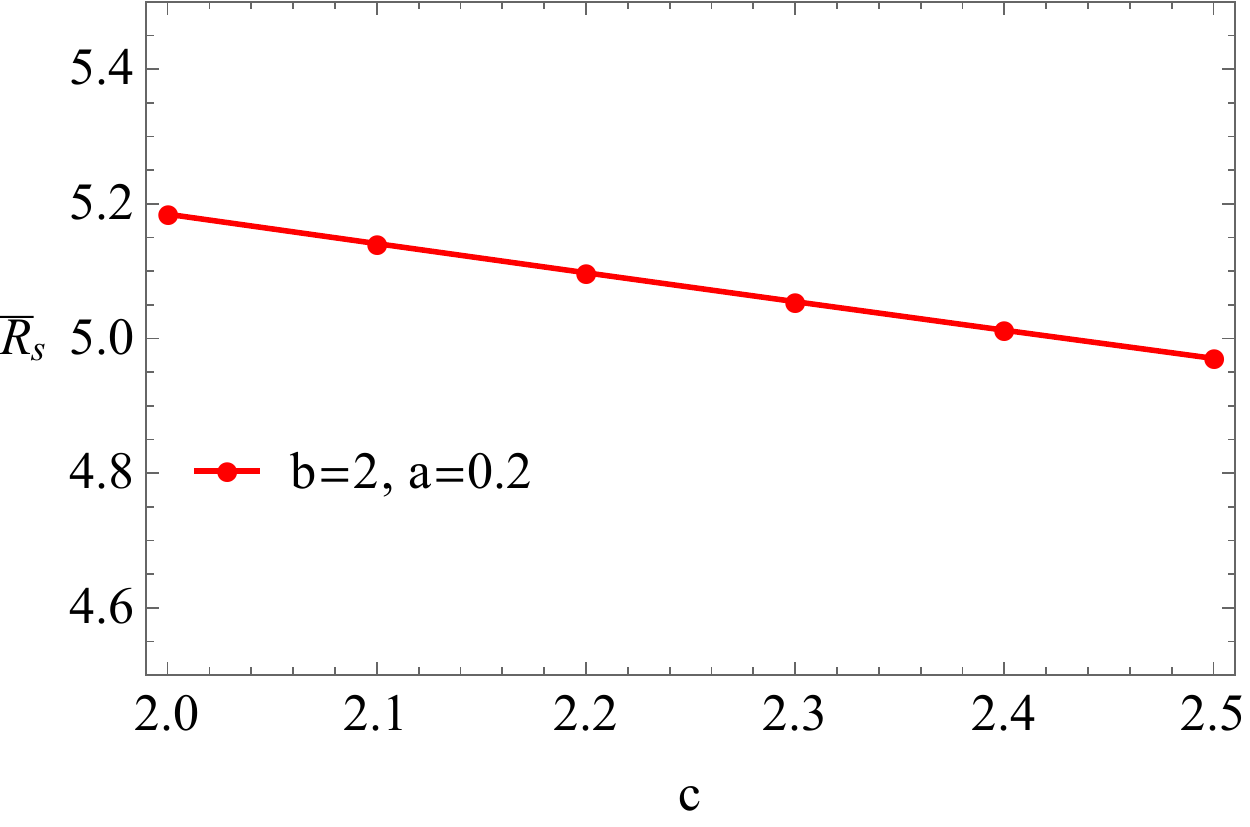}
\includegraphics[width=7.4cm]{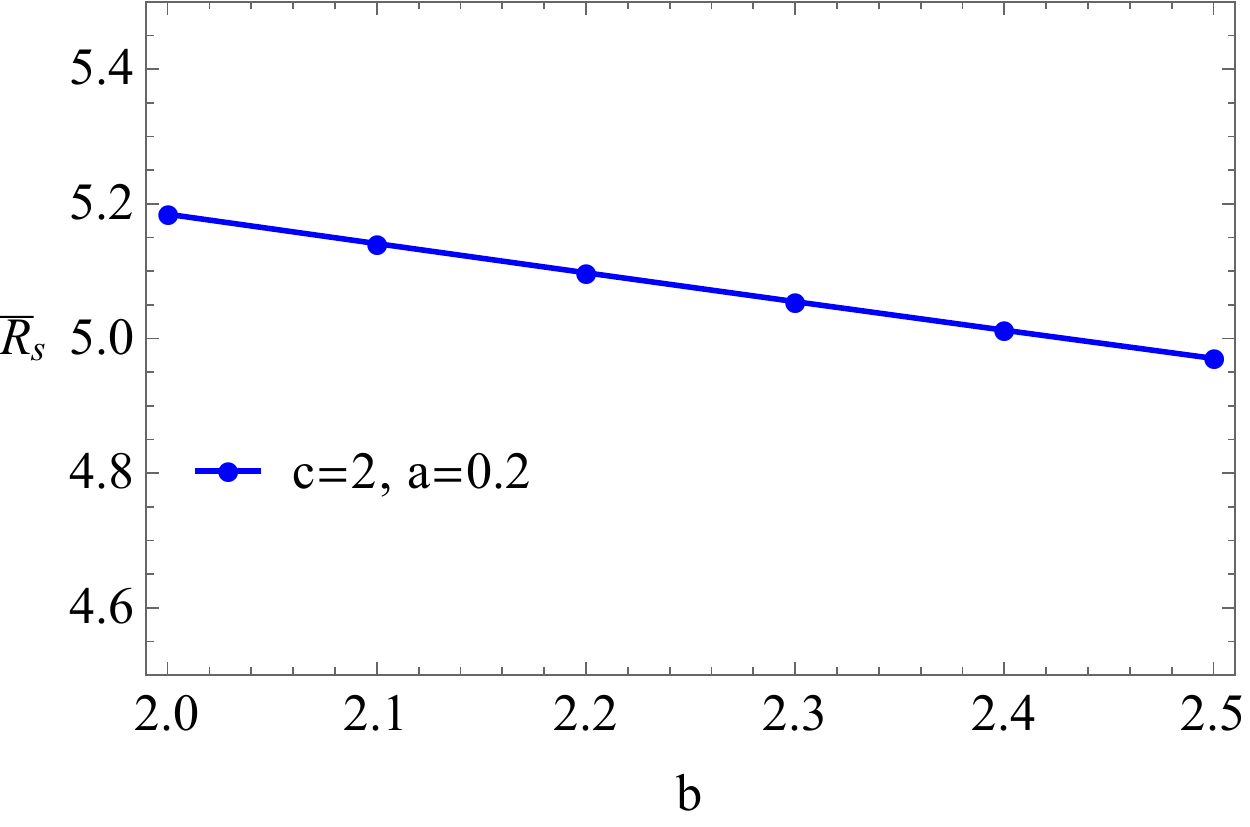}
\caption{Left panel: Variation of the typical shadow radius for KK black hole as a function of $c$. Right panel: Variation of the typical shadow radius for KK black hole as a function of $b$.  The inclination angle is $i=\pi/2$. }\label{FigR}
\end{figure*}

Then from the last two equations yields
\begin{equation}
g'_{\phi \phi}|_{r_0}+2 g'_{t \phi}|_{r_0} R_s+g'_{tt}|_{r_0} R_s^2=0,
\end{equation}
which has the following solutions
\begin{equation}\label{Eq34}
R_s^{\pm}=-\frac{g'_{t \phi}|_{r_0} \pm \sqrt{(g'_{t \phi})^2|_{r_0}-g'_{\phi \phi}|_{r_0}\, g'_{t t}|_{r_0}}}{g'_{t t}|_{r_0}}.
\end{equation}

If we now combine this result with Eq. \eqref{Eq30} we obtain
\begin{equation}\label{Eq35}
g_{\phi \phi}|_{r_0^{\pm}}+2 g_{t \phi}|_{r_0^{\pm}} R_s^{\pm} +g_{t t}|_{r_0^{\pm}} (R_s^{\pm})^2=0.
\end{equation}

From the last equation  we can determine the point $r=r_0^{\pm}$, which gives the radius of circular null geodesics for the retrograde and prograde case, respectively.
In order to compute the shadow radius we point out that, in general, the shape of the shadow depends on the observer's viewing angle $i$, however in the case with $i=\pi/2$, we can adopt the definition known as the typical shadow radius which can be written as \cite{Jusufi:2020dhz,Feng:2019zzn}
\begin{equation}
\bar{R}_s=\frac{1}{2}\left(R^+_s|_{r_0^+}-R^-_s|_{r_0^-}\right)
\end{equation}
where $r_0^{\pm}$ is determined from Eq. \eqref{Eq35}. The final result can be written as
\begin{widetext}
\begin{eqnarray}\label{Eq37}
\bar{R}_s&=&\frac{1}{2}\Big(\frac{g'_{t \phi}|_{r_0^-}}{g'_{t t}|_{r_0^-}}-\frac{g'_{t \phi}|_{r_0^+}}{g'_{t t}|_{r_0^+}}\Big)- \frac{1}{2}\left(\frac{\sqrt{(g'_{t \phi})^2|_{r_0^+}-g'_{\phi \phi}|_{r_0^+}\, g'_{t t}|_{r_0^+}}}{g'_{t t}|_{r_0^+}}+\frac{\sqrt{(g'_{t \phi})^2|_{r_0^-}-g'_{\phi \phi}|_{r_0^-}\, g'_{t t}|_{r_0^-}}}{g'_{t t}|_{r_0^-}}\right).
\end{eqnarray}
\end{widetext}

The last equation generalizes the results obtained in Ref. \cite{Jusufi:2020dhz}. As a special case setting $b=c=2$ we can obtain the typical shadow radius for the Kerr spacetime.  To see this let us write the metric components for the Kerr black hole
\begin{eqnarray}
g_{t\phi}&=&-a (1-f(r)),\\
g_{tt}&=&-f(r),\\
g_{\phi \phi}&=& [(r^2+a^2)^2-a^2\Delta]/r^2,
\end{eqnarray}
where
\begin{equation}
\Delta=r^2 f(r)+a^2,\,\,\,f(r)=1-\frac{2M}{r}.
\end{equation}

From Eq. \eqref{Eq37} we therefore obtain
\begin{equation}
\bar{R}_s=\frac{\sqrt{2}}{2}\left(\sqrt{\frac{ r_0^{+}}{f'(r)|_{r_0^{+}}}}+\sqrt{\frac{ r_0^{-}}{f'(r)|_{r_0^{-}}}}\right),
\end{equation}
which has been reported in Ref. \cite{Jusufi:2020dhz}.
In what follows, we shall specialize our discussion for the KK black hole having  $b=2$ along with $M=1$, yielding the metric components
\begin{equation}
g_{tt}=-\frac{r-\frac{8}{2+c}}{\sqrt{r^2+\frac{4 (c-2) r}{c+2}}},
\end{equation}
\begin{equation}
g_{t \phi}=-\frac{4 \sqrt{2 \,c} \, a }{(2+c)\sqrt{r^2+\frac{4 (c-2) r}{c+2}}},
\end{equation}
\begin{equation}
g_{\phi \phi }=\frac{(r+4)(r^2+a^2)\,c+2 r a^2+2 r^3-8 r^2}{\sqrt{r\, (2+c)\,\left( (r+4)c+2 r-8  \right)}},
\end{equation}
having two solutions for $R^{\pm}_s$, given by
\begin{eqnarray}
R^{\pm}_s=\frac{(2 a r +4 a)\sqrt{2\,c}\, c+(4a r -8a) \sqrt{2  \,c} \pm \sqrt{\zeta}}{(c+2)^2 r+8c -16}.
\end{eqnarray}
where
\begin{eqnarray}
\zeta &= & r \left((c+2) r +4 c-8\right)\,\\\notag
& \times & \Big[(c+2)^3 r^3-r^2 \mathcal{A}+24 (c-2)^2 r -a^2 (c+2)(c-2)^2\Big]
\end{eqnarray}
and \begin{equation}
\mathcal{A}=-3 c^3-14 c^2+12 c+56.
\end{equation}

If we use the geometric-optics correspondence between the parameters of a quasinormal mode, and the conserved quantities
along geodesics we can identify the energy of the particle with the real part of QNMs, hence
\begin{equation}
E \to \omega_{\Re},
\end{equation}
while the azimuthal quantum number corresponds to angular momentum
\begin{equation}
J \to \mathrm{m}.
\end{equation}

In the eikonal limit in the rotating spacetimes we have
\begin{equation}
\mathrm{m}=\pm l,
\end{equation}
corresponding to the prograde and retrograde modes, respectively. With these identifications and following the arguments in Refs. \cite{Jusufi:2019ltj,Jusufi:2020dhz} we can  write
\begin{equation}
\omega_{\Re}^{\pm} = \lim_{l \gg 1}  \frac{\mathrm{m}}{R_s^{\pm}}.
\end{equation}

If we combine this relation with the result in Eq. \eqref{Eq34}, we obtain the real part of QNMs
\begin{equation}
\omega_{\Re}^{\pm} = \lim_{l \gg 1} \left[ \frac{\mathrm{m}}{-\frac{g'_{t \phi}|_{r_0^{\pm}} \pm \sqrt{(g'_{t \phi})^2|_{r_0^{\pm}}-g'_{\phi \phi}|_{r_0^{\pm}}\, g'_{t t}|_{r_0^{\pm}}}}{g'_{t t}|_{r_0^{\pm}}}}\right].
\end{equation}

\begin{table}[tbp]
\begin{tabular}{|l|l|l|l|l|l|}
\hline
\multicolumn{1}{|c|}{ } &  \multicolumn{1}{c|}{  $\mathrm{m}=100$ } & \multicolumn{1}{c|}{  $\mathrm{m}=100$ } & \multicolumn{1}{c|}{KK}\\\hline
  $a/M$ &\,\,\,\,$\omega_{\Re}^{+}$  &\,\,\,\,$\omega_{\Re}^{-}$  & \,\,\,\,$\bar{R}_s$   \\ \hline
0.0 & 19.406207 & -19.406207   & 5.152990 \\
0.1 & 20.209522 & -18.685364   & 5.149972 \\
0.2 & 21.113973 & -18.033018  & 5.140791  \\
0.3 & 22.145295 & -17.438429 &  5.125046  \\
0.4 & 23.340431 & -16.893166 &  5.101929 \\
0.5 & 24.755291 & -16.390489 &  5.070319  \\\hline
\end{tabular}
\caption{The typical shadow radius for the KK black hole by varying the angular momentum parameter $a$ with fixed $b=2$ and $c=2.1$. Note that the result are similar if we chose $c=2$ and $b=2.1$.}
\end{table}

\begin{figure*}
\includegraphics[width=7.4cm]{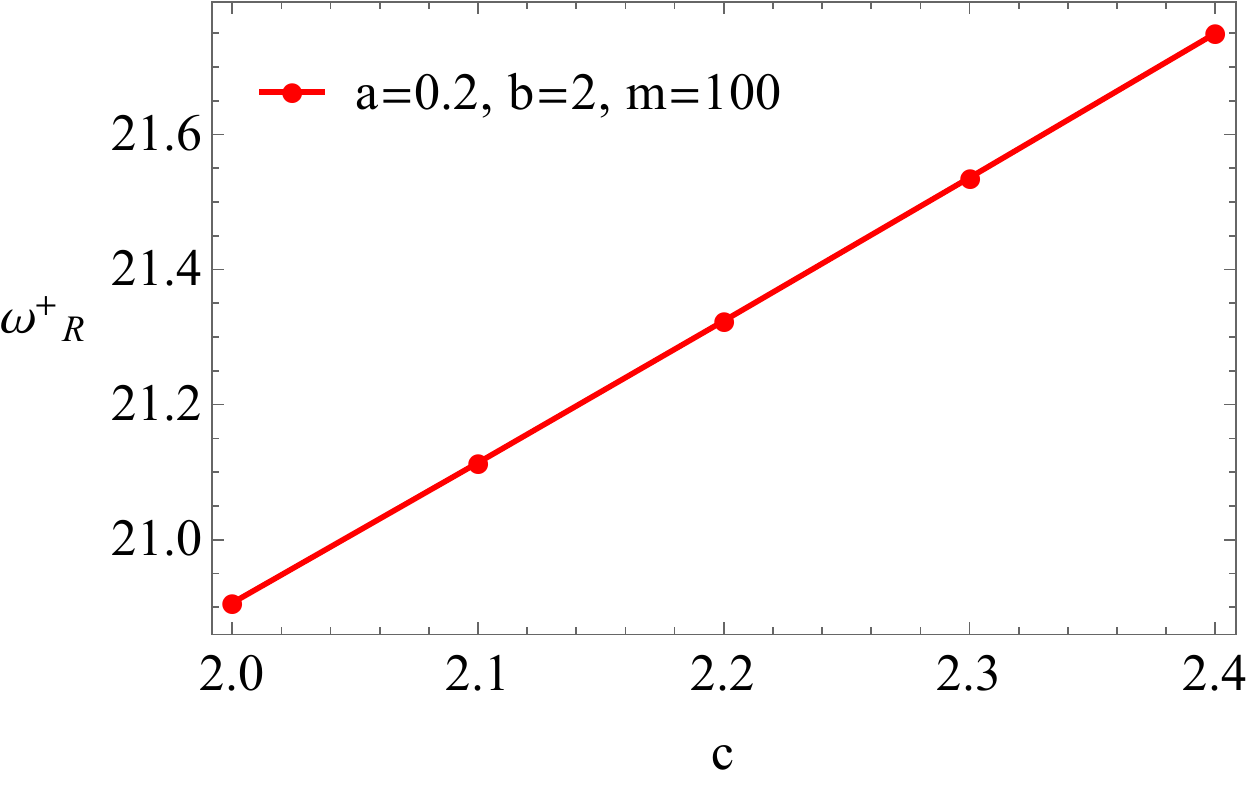}
\includegraphics[width=7.4cm]{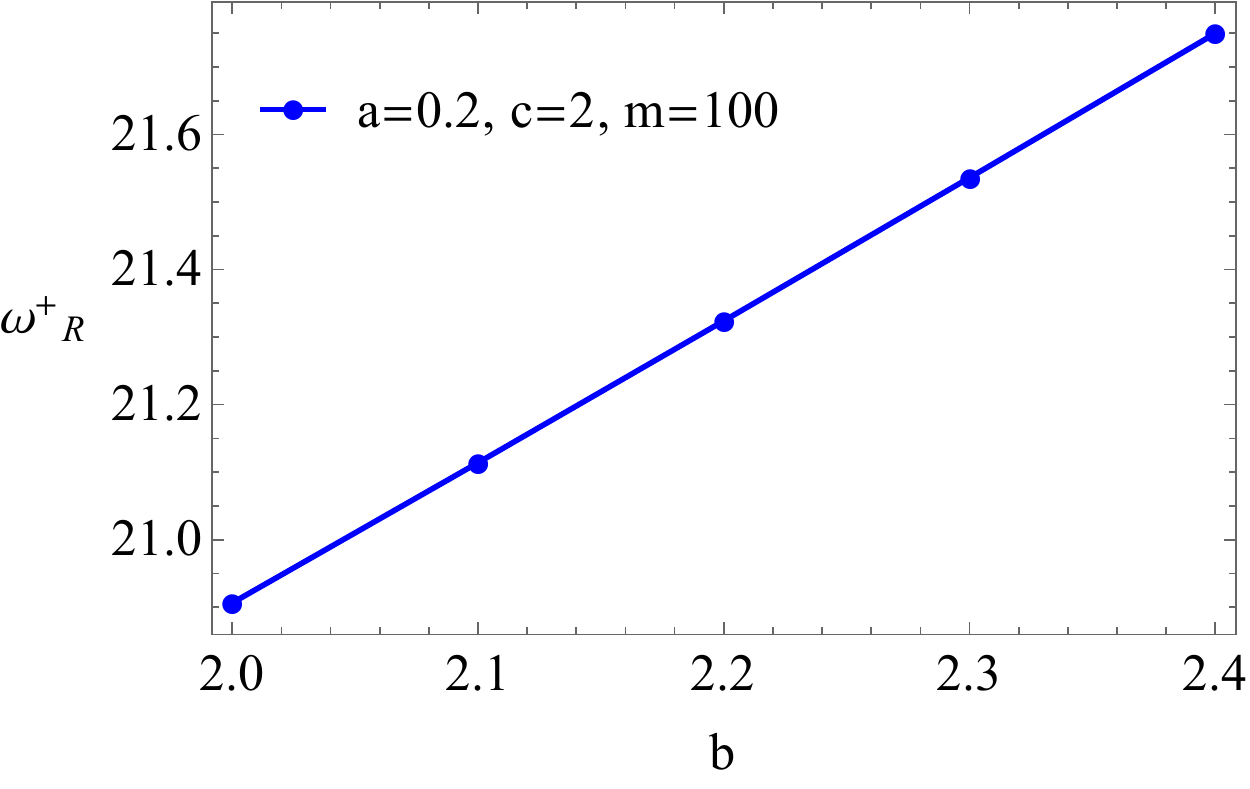}
\caption{Left panel: Variation of the positive branch of the ral part of QNMs  as a function of $c$. Right panel:  Variation of the positive branch of the ral part of QNMs  as a function of $b$.}\label{FigW}
\end{figure*}

Finally, let us define the following quantity for the real part of QNMs
\begin{equation}
\bar{\omega}_{\Re}=\frac{1}{2}(\omega_{\Re}^+-\omega_{\Re}^-),
\end{equation}
in terms of this equation, we can write the following correspondence
\begin{equation}\label{Eq52}
\bar{\omega}_{\Re} = \lim_{l \gg 1}  \frac{\mathrm{m}}{\bar{R}}_s.
\end{equation}

From Table I we see that for a fixed values of $b$ or $c$, the typical shadow radius of the rotating KK black hole decreases with the increase of the angular momentum parameter. On the other hand, from Fig. 8 we observe that for e fixed angular momentum parameter, the typical shadow radius decreases with the increase of $c$ and $b$, respectively. Note that, the typical shadow radius of KK black hole is smaller compared to the Kerr black hole in the domain of parameters chosen in Fig. \ref{FigR}. Furthermore, the positive branch of the real part of QNMs increases with the increase of $c$ and $b$, respectively (see Fig. \ref{FigW}). Since the typical shadow radius decreases with the increase of $c$ and $b$, we can see that the value of $\bar{\omega}_{\Re}$ increases. In fact, this can be understood easily due to the inverse relation between $\bar{\omega}_{\Re}$ and the typical shadow radius given by Eq. \eqref{Eq52}.\\

\section{Quasi-periodic oscillations (QPOs)\label{secqpos}}

The observational constraints on the parameters of the four dimensional Kaluza-Klein black hole could also be determined via theoretical calculations of the quasi-periodic oscillations (QPOs) and curve fitting to existing experimental data for rotating microquasars. The appearance of two peaks at 300 Hz and 450 Hz in the X-ray power density spectra of Galactic microquasars, representing possible occurrence of a lower QPO and of an upper QPO in a ratio of 3 to 2, has stimulated a lot of theoretical works to explain the value of the 3/2-ratio. In this paper we rely on the parametric resonance model for our investigations and we describe a microquasar as a four dimensional Kaluza-Klein black hole. For that purpose, we have selected three microquasars the astrophysical data of which are the most accurate.

In this section we need the numerical values of some physical constants including the solar mass $M_\odot=1.9888\times 10^{30}$, the gravitational constant $G=6.673\times 10^{-11}$, and the speed of light in vacuum $c_{\text{v}}=299792458$, all given in SI units. These same constants will be written explicitly in some subsequent formulas of this section.

The power spectra of Fig.~3 of Ref.~\cite{res} clearly reveal two peaks at 300 Hz and 450 Hz  representing, respectively, the possible occurrence of the lower $\nu_L=300$ Hz quasi-periodic oscillation (QPO), and of the upper $\nu_U=450$ Hz QPO from the Galactic microquasar GRO J1655-40. Similar peaks have been obtained for the microquasars XTE J1550-564 and GRS 1915+105 obeying the remarkable
relation,  $\nu_U/\nu_L=3/2$~\cite{qpos1}. The relevant physical quantities of these three microquasars and their uncertainties are as follows~\cite{res,res2}:
\begin{multline}\label{pr1}
\text{GRO J1655-40 : }\frac{M}{M_\odot}=6.30\pm 0.27,\;\frac{a}{r_g}=0.70\pm 0.05\\
\nu_U=450\pm 3 \text{ Hz},\;\nu_L=300\pm 5 \text{ Hz},
\end{multline}
\begin{multline}\label{pr2}
\text{XTE J1550-564 : }\frac{M}{M_\odot}=9.1\pm 0.6,\;\frac{a}{r_g}=0.405\pm 0.115\\
\nu_U=276\pm 3 \text{ Hz},\;\nu_L=184\pm 5 \text{ Hz},
\end{multline}
\begin{multline}\label{pr3}
\text{GRS 1915+105 : }\frac{M}{M_\odot}=14.0\pm 4.4,\;\frac{a}{r_g}=0.99\pm 0.01\\
\nu_U=168\pm 3 \text{ Hz},\;\nu_L=113\pm 5 \text{ Hz},
\end{multline}
where $r_g\equiv GM/c^2$.

These twin values of the QPOs are most certainly due to the phenomenon of resonance which occurs in the vicinity of the ISCO, where the in-falling  particles perform radial and vertical oscillations around almost circular orbits. These two oscillations couple generally non-linearly to yield resonances in the power spectra~\cite{res3,res4}.

In the first part of this section we will be concerned with stable circular orbits in the plane of symmetry (the plane $\theta=\pi/2$) and their perturbations, since these orbits constitute the trajectories of in-falling matter in accretion processes.

From now on we consider stable circular orbits of uncharged particles in the $\theta=\pi/2$ plane. First of all, we need to set up the equations governing an unperturbed circular motion. Once this is done, we will derive the equations that describe a perturbed circular motion around a stable unperturbed circular motion. In a third step we will {separate out} the set of equations governing the perturbed circular motion.

The unperturbed circular motion is a geodesic motion obeying the equation,
\begin{equation}\label{p1}
\frac{\dee u^{\mu}}{\dee \tau}+\Gamma^{\mu}_{\alpha\beta}u^{\alpha}u^{\beta}=
0,
\end{equation}
where $u^{\mu}=\dee x^{\mu}/\dee \tau =\dot x^{\mu}$ is the four-velocity. Here the connection $\Gamma^{\mu}_{\alpha\beta}$ is related to the unperturbed metric~\eqref{eq-4dmetric}. For a circular motion in the equatorial plane ($\theta=\pi/2$), $u^{\mu}=(u^t,\,0,\,0,\,u^{\phi})=u^t(1,\,0,\,0,\,\omega)$, where $\omega=\dee\phi/\dee t$ is the angular velocity of the test particle. The only equation describing such a motion is the $r$ component of~\eqref{p1} and the normalization condition $g_{\mu\nu}u^{\mu}u^{\nu}=-c^2$, which take,  respectively, the following forms,
\begin{align}
\label{p2a}&\partial _r g_{tt} (u^t)^2+2 \partial _r g_{t\phi}u^t u^{\phi }+\partial _r g_{\phi \phi } (u^{\phi })^2=0,\\
\label{p2b}&g_{tt} (u^t)^2+2 g_{t\phi} u^t u^{\phi }
+g_{\phi \phi } (u^{\phi })^2=-c^2,
\end{align}
where the metric and its derivatives are evaluated at $\theta=\pi/2$. Solving~\eqref{p2a} and~\eqref{p2b}, we  obtain
\begin{align}
&\omega =\frac{-\partial _rg_{t \phi }\pm \sqrt{\left(\partial _rg_{t \phi }\right)^2-\partial _rg_{t t} \partial _rg_{\phi  \phi }}}{\partial _rg_{\phi
		\phi }},\nonumber\\
&u^t=\frac{c}{\sqrt{-\left(g_{t t}+2 \partial _rg_{t \phi } \omega +g_{\phi  \phi } \omega ^2\right)}},\nonumber\\
\label{p3}&u^{\phi }=\omega  u^t ,
\end{align}
where the upper sign corresponds to prograde circular orbits and the lower sign corresponds to retrograde orbits.

If the motion is perturbed, the actual position is now denoted by $X^{\mu}=x^{\mu}+\eta^{\mu}$ and the 4-velocity by $U^{\mu}=u^{\mu}+\dot{\eta}^{\mu}$ (where $~\dot{}\equiv \dee /\dee\tau$) with $u^{\mu}$ being the unperturbed values given in~\eqref{p3}. First substituting $U^{\mu}=u^{\mu}+\dot{\eta}^{\mu}$ into
\begin{equation}
\frac{\dee U^{\mu}}{\dee \tau}+\Gamma^{\mu}_{\alpha\beta}(X^{\sigma})U^{\alpha}U^{\beta}=
0,
\end{equation}
where $\Gamma^{\mu}_{\alpha\beta}(X^{\sigma})$ is the perturbed connection, { and then keeping only linear terms in $\eta^\mu$ and its derivatives
	(and also considering  \eqref{p1}), we finally arrive at~\cite{Kerr1,qposknb}
	\begin{equation}\label{p4}
	\ddot{\eta}^{\mu}+2\Gamma^{\mu}_{\alpha\beta}u^{\alpha}\dot{\eta}^{\beta}
	+\partial_{\nu}\Gamma^{\mu}_{\alpha\beta}u^{\alpha}u^{\beta}\eta^{\nu}
	=0,
	\end{equation}
	where the background connection $\Gamma^{\mu}_{\alpha\beta}$ and its derivatives are  all evaluated at $\theta=\pi/2$.} As shown in~\cite{qposknb}, Eqs.~\eqref{p4} decouple and take the form of oscillating radial (in the $\theta=\pi/2$ plane) and vertical (perpendicular to the $\theta=\pi/2$ plane) motions obeying the following harmonic equations:
\begin{align}
&\ddot{\eta}^{r}+\Omega_{r}^2\eta^{r}=0,
&\ddot{\eta}^{\theta}+\Omega_{\theta}^2\eta^{\theta}=0.
\end{align}
The locally measured frequencies ($\Omega_{r},\,\Omega_{\theta}$) are related to the spatially-remote observer's frequencies ($\nu_{r},\,\nu_{\theta}$) by
\begin{align}\label{pr4}
&\nu_r=\frac{1}{2\pi}~\frac{1}{u^t}~\Omega_{r},
&\nu_\theta=\frac{1}{2\pi}~\frac{1}{u^t}~\Omega_{\theta},
\end{align}
where $u^t$ is given in~\eqref{p3} and~\cite{qposknb}
\begin{align}
&\Omega_{\theta}^2\equiv (\partial_{\theta}\Gamma^{\theta}_{ij})u^iu^j,\qquad (i,\,j=t,\,\phi),\\
\label{p5b}&\Omega_{r}^2\equiv (\partial_{r}\Gamma^{r}_{ij}-4\Gamma^{r}_{ik}\Gamma^{k}_{rj})u^iu^j,\qquad (i,\,j,\,k=t,\,\phi).
\end{align}
In these expressions the summations extend over ($t,\,\phi$). It is understood that all the functions appearing in~\eqref{p3}, \eqref{pr4} and~\eqref{p5b} are evaluated at $\theta=\pi/2$.

Next, we introduce the dimensionless parameters $y$ and $a_0$ defined by
\begin{equation}\label{rg}
y\equiv\frac{r}{r_g},\quad a_0\equiv\frac{a}{r_g},\quad r_g\equiv\frac{GM}{c^2},
\end{equation}
in terms of which we determine the expressions of ($\nu_{r},\,\nu_{\theta}$) measured in Hz. However, these expression are so lengthy and cannot be given here.

As we mentioned earlier, the twin values of the QPOs observed in the microquasars are most certainly due to the phenomenon of resonance resulting from the coupling of the vertical and radial oscillatory motions~\cite{res3,res4}. The most common models for resonances are parametric resonance, forced resonance and Keplerian resonance. {It is the general belief} that the resonance observed in the three microquasars~\eqref{pr1}, \eqref{pr2} and~\eqref{pr3} is  {of the nature of the parametric resonance and is given by}
\begin{equation}\label{as1}
\nu_U = \nu_\theta,\qquad \nu_L =\nu_r \ ,
\end{equation}
with
\begin{equation}\label{as2}
\frac{\nu_\theta}{\nu_r}=\frac{n}{2} ,\qquad n\in \mathbb{N}^+ .
\end{equation}
{In most of the applications of the} parametric resonance one considers the case $n=1$~\cite{b1,b2,b3,b4}, where in this case $\nu _{r}$ is the natural frequency of the system and $\nu_{\theta }$ is the parametric excitation ($T_{\theta }=2T_{r}$, the corresponding periods), that is, the vertical oscillations supply energy to the radial oscillations causing resonance~\cite{b4}. However, since $\nu_\theta>\nu_r$ in the vicinity of ISCO, where accretion occurs and QPO resonance effects take place, the lower possible value of $n$ is 3 and in this case $\nu_{r}$ becomes the parametric excitation that supplies energy to the vertical oscillations.

Thus, the observed ratio $\nu_U/\nu_L=3/2$ is theoretically justified by making the assumptions~\eqref{as1} and~\eqref{as2} with $n=3$. Numerically we have to show that the plot of $\nu_U=\nu_\theta$ ($\nu_L=\nu_r$) versus $M/M_\odot$ crosses the upper (lower) mass band error, given in~\eqref{pr1}, \eqref{pr2} and~\eqref{pr3}, as $a_0$ assumes values in its defined band error and this will allow us to determine the range of the dimensionless parameter $c$ for each microquasar. Since magnetic charges have never been observed we take $b=2$ yielding $p=2m$ and $P=0$~\eqref{eq-rels}. Introducing the relevant universal constants and using~\eqref{eq-rels} we arrive at
\begin{equation}\label{ratio}
\frac{Q^2}{4\pi\epsilon_0 G M^2}=\frac{4c(c-2)}{(c+2)^2}.
\end{equation}
Since the function in the right-hand side of~\eqref{ratio} is always increasing for $c\geq 2$, the numerical constrained values of $c$, $c_{\text{min}}\leq c\leq c_{\text{max}}$, yield the following limits on the ratio $Q^2/M^2$
\begin{equation}\label{ratio2}
0\leq \frac{4c_{\text{min}}(c_{\text{min}}-2)}{(c_{\text{min}}+2)^2}\leq \frac{Q^2}{4\pi\epsilon_0 G M^2}\leq \frac{4c_{\text{max}}(c_{\text{max}}-2)}{(c_{\text{max}}+2)^2}<4.
\end{equation}

\begin{figure*}
	\includegraphics[width=7.4cm]{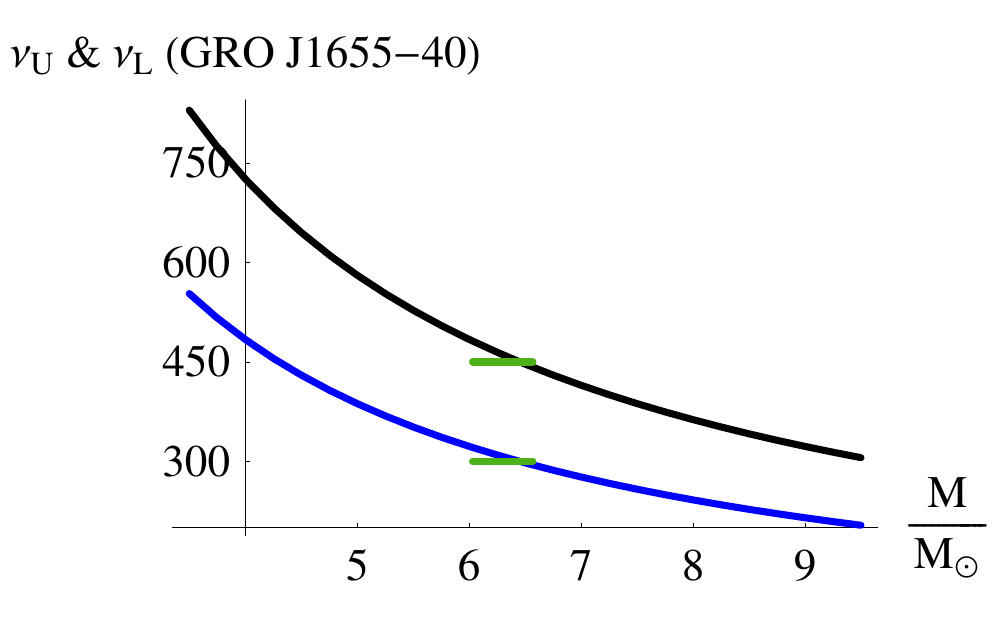}
	\includegraphics[width=7.4cm]{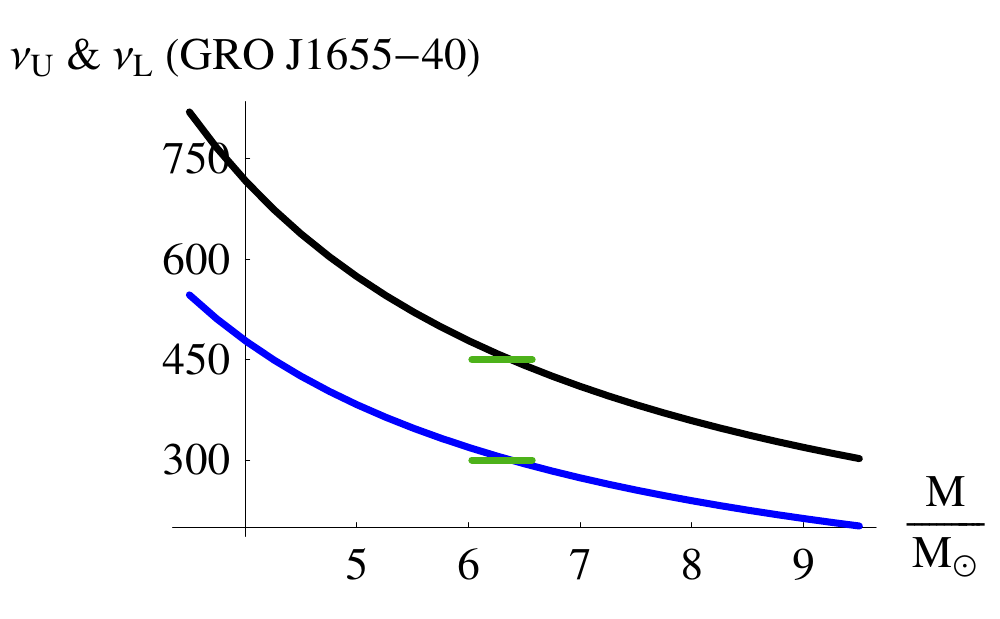}
	\caption{The black (blue) curve is a plot of $\nu_U=\nu_\theta$ ($\nu_L=\nu_r$) in Hz versus $M/M_\odot$ for the microquasar GRO J1655-40 treated as a rotating Kaluza-Klein BH taking $P=0$ ($b=2$). The green curves represent the mass limits as given in~\eqref{pr1}. Left panel: The location of the outer horizon and the smallest root of~\eqref{as2} ($n=3$) are determined taking $M/M_\odot =6.5$ and $a/r_g=0.65$. Right panel: The location of the outer horizon and the smallest root of~\eqref{as2} ($n=3$) are determined taking $M/M_\odot =6.1$ and $a/r_g=0.75$. This curve fitting yields the limits $3.20\leq c\leq 3.95$.}\label{Figgro}
\end{figure*}
\begin{figure*}
	\includegraphics[width=7.4cm]{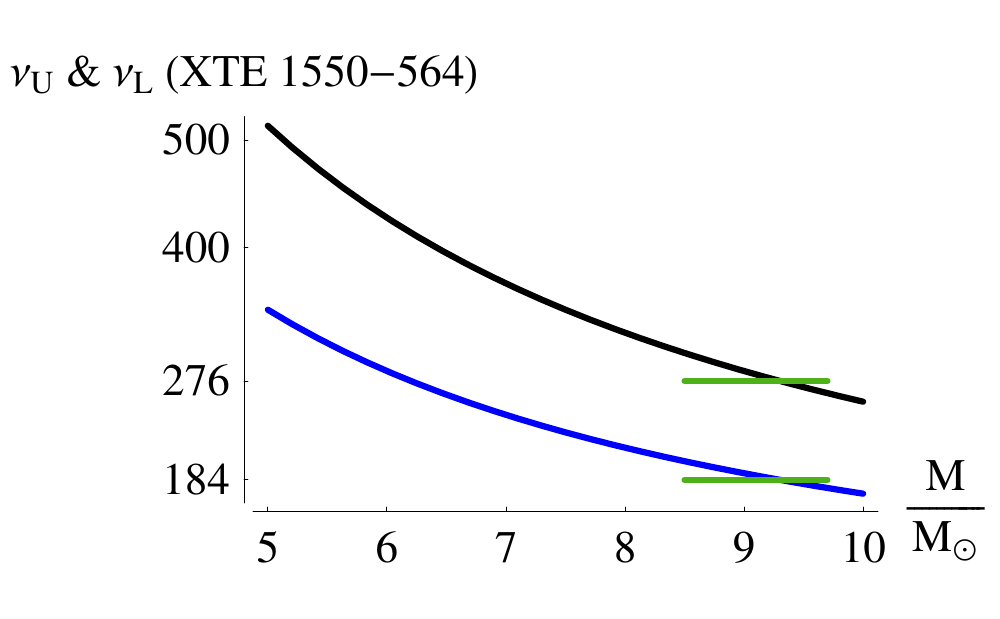}
	\includegraphics[width=7.4cm]{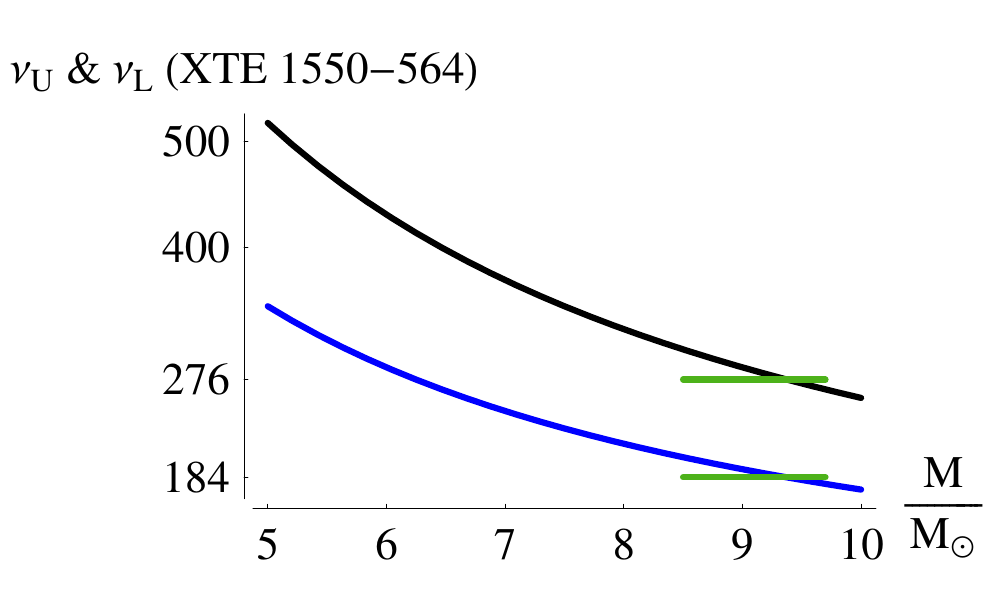}
	\caption{The black (blue) curve is a plot of $\nu_U=\nu_\theta$ ($\nu_L=\nu_r$) in Hz versus $M/M_\odot$ for the microquasar XTE J1550-564 treated as a rotating Kaluza-Klein BH taking $P=0$ ($b=2$). The green curves represent the mass limits as given in~\eqref{pr2}. Left panel: The location of the outer horizon and the smallest root of~\eqref{as2} ($n=3$) are determined taking $M/M_\odot =9.7$ and $a/r_g=0.29$. Right panel: The location of the outer horizon and the smallest root of~\eqref{as2} ($n=3$) are determined taking $M/M_\odot =8.5$ and $a/r_g=0.52$. This curve fitting yields the limits $4.70\leq c\leq 7.50$.}\label{Figxte}
\end{figure*}
\begin{figure*}
	\includegraphics[width=7.4cm]{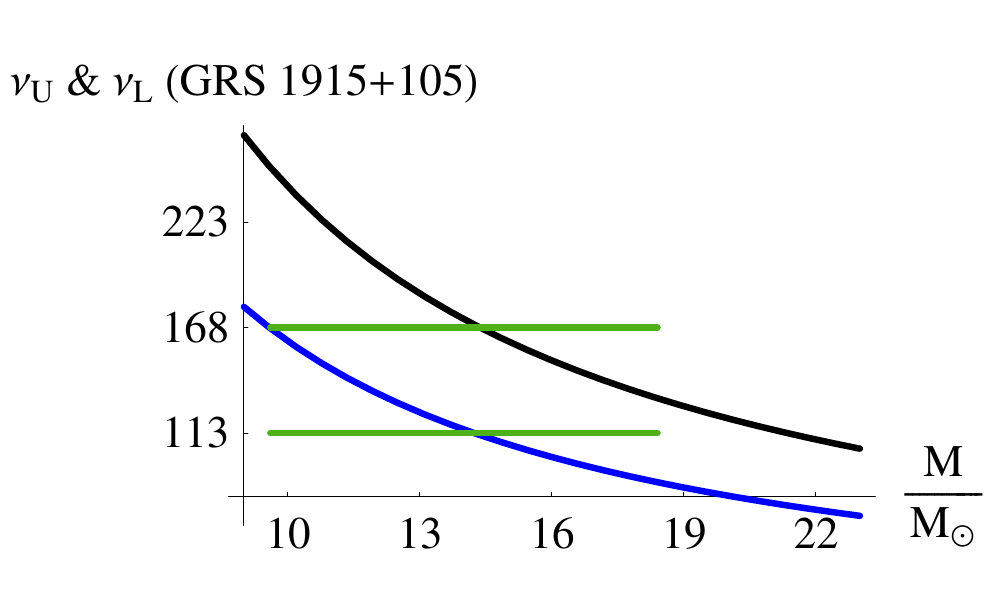}
	\includegraphics[width=7.4cm]{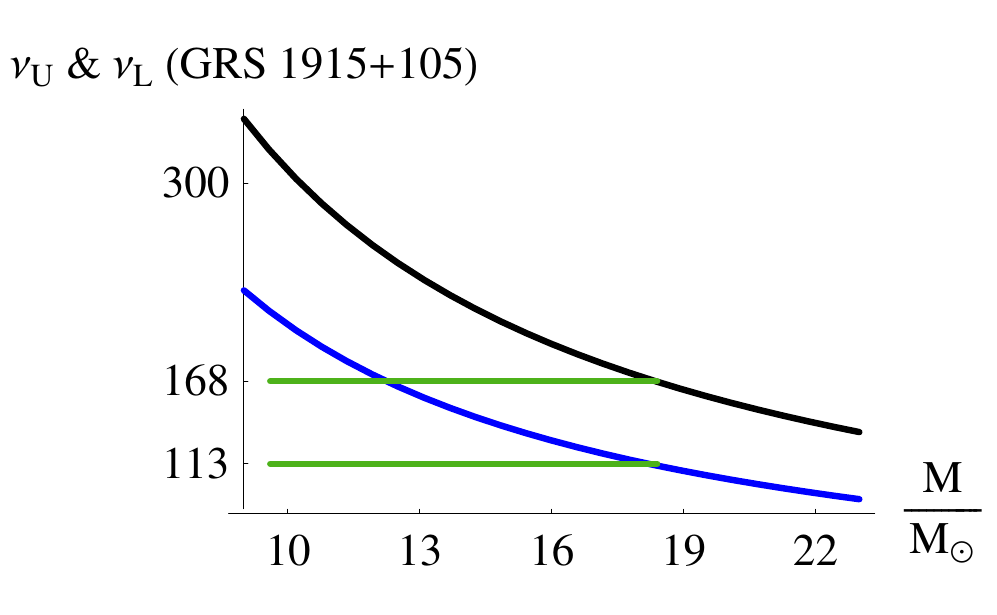}
	\caption{The black (blue) curve is a plot of $\nu_U=\nu_\theta$ ($\nu_L=\nu_r$) in Hz versus $M/M_\odot$ for the microquasar GRS 1915+105 treated as a rotating Kaluza-Klein BH taking $P=0$ ($b=2$). The green curves represent the mass limits as given in~eqref{pr3}. Left panel: The location of the outer horizon and the smallest root of~\eqref{as2} ($n=3$) are determined taking $M/M_\odot =18.4$ and $a/r_g=0.98$. Right panel: The location of the outer horizon and the smallest root of~\eqref{as2} ($n=3$) are determined taking $M/M_\odot =9.6$ and $a/r_g=1.00$. This curve fitting yields the limits $2.00\leq c\leq 2.06$.}\label{Figgrs}
\end{figure*}
\begin{figure}
	\includegraphics[width=7.4cm]{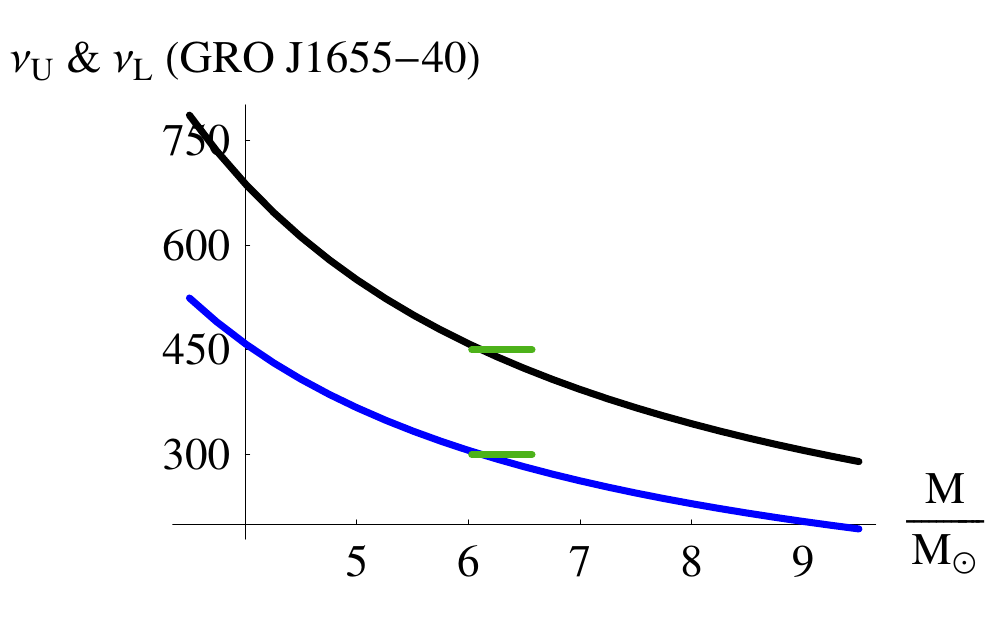}
	\caption{The black (blue) curve is a plot of $\nu_U=\nu_\theta$ ($\nu_L=\nu_r$) in Hz versus $M/M_\odot$ for the microquasar GRO J1655-40 treated as a Kerr--Newman BH. The green curves represent the mass limits as given in~\eqref{pr1}. This curve fitting is possible only for $Q^2/(4\pi\epsilon_0 GM^2)\sim 0.66$.}\label{Figkngro}
\end{figure}

Curves that fit the  upper and lower oscillation frequencies  of the uncharged test particles to the observed frequencies (in Hz) of the microquasars GRO J1655-40, XTE J1550-564 and GRS 1915+105 at the 3/2 resonance radius are presented in Figs.~\ref{Figgro}, \ref{Figxte} and~\ref{Figgrs}, respectively. In these plots each microquasar is treated as a rotating Kaluza-Klein BH~\eqref{eq-4dmetric} with $P=0$ ($b=2$). The black curves represent $\nu_U=\nu_\theta$ versus $M/M_\odot$, the blue curves represent $\nu_L=\nu_r$ versus $M/M_\odot$ with $\nu_U/\nu_L=3/2$, and the green curves represent the mass limits as given in~\eqref{pr1}, \eqref{pr2} and~\eqref{pr3}. As indicated in the caption of each figure, the location of the outer horizon and the smallest root of~\eqref{as2} ($n=3$) in the left panel are determined taking the largest value of $M/M_\odot$ (keeping one decimal place after the comma) and to the smallest value of $a/r_g$, and the location of the outer horizon and the smallest root of~\eqref{as2} ($n=3$) in the right panel are determined taking the smallest value of $M/M_\odot$ (keeping one decimal place after the comma) and to the largest value of $a/r_g$.

\subsection{Microquasar GRO J1655-40}
In the left panel of Fig~\ref{Figgro} in order to determine the location of the outer horizon and the smallest root of~\eqref{as2} ($n=3$) we assumed the standard mass and the rotation parameter of the Kaluza-Klein BH~\eqref{eq-4dmetric} to be $M/M_\odot =6.5$ and $a/r_g=0.65$, respectively, then we varied the mass within the limits~\eqref{pr1} to obtain plots of $\nu_U$ and $\nu_L$. This yields the numerical upper limit $c_{\text{max}}=3.95$. For the right panel of Fig~\ref{Figgro} we assumed the standard mass and the rotation parameter of the Kaluza-Klein BH to be $M/M_\odot =6.1$ and $a/r_g=0.75$, respectively. This yields the numerical lower limit $c_{\text{min}}=3.20$. Using~\eqref{ratio2} we obtain
\begin{equation}\label{ratio3}
0.57\lesssim \frac{Q^2}{4\pi\epsilon_0 G M^2}\lesssim 0.87,
\end{equation}
which is within classical limits.

Had we treated this microquasar as a Kerr-Newman BH, as in Fig.~\ref{Figkngro}, the curve fitting would possible only for $Q^2/(4\pi\epsilon_0 GM^2)\sim 0.66$, which is well within the limits given in~\eqref{ratio3}. We see that Kaluza-Klein BHs allow for a set of possible values for the ratio $Q^2/(4\pi\epsilon_0 GM^2)$.

\subsection{Microquasar XTE J1550-564}
In the left panel of Fig~\ref{Figxte} in order to determine the location of the outer horizon and the smallest root of~\eqref{as2} ($n=3$) we assumed the standard mass and the rotation parameter of the Kaluza-Klein BH to be $M/M_\odot =9.7$ and $a/r_g=0.29$, respectively, then we varied the mass within the limits~\eqref{pr2} to obtain plots of $\nu_U$ and $\nu_L$. This yields the numerical upper limit $c_{\text{max}}=7.50$. For the right panel of Fig~\ref{Figgro} we assumed the standard mass and the rotation parameter of the Kaluza-Klein BH to be $M/M_\odot =8.5$ and $a/r_g=0.52$, respectively. This yields the numerical lower limit $c_{\text{min}}=4.70$. Using~\eqref{ratio2} we obtain
\begin{equation}\label{ratio4}
1.13\lesssim \frac{Q^2}{4\pi\epsilon_0 G M^2}\lesssim 1.83.
\end{equation}
This shows that for this microquasar to be treated as a Kaluza-Klein BH the ratio $Q^2/(4\pi\epsilon_0 G M^2)$ should be greater than unity.

There would be no room to obtain a satisfactory curve fitting had we treated this microquasar as a Kerr-Newman BH; the curve fitting would yield the empty set $\Phi$ for the possible values of $Q^2/(4\pi\epsilon_0 G M^2)$.

\subsection{Microquasar GRS 1915+105}
In the left panel of Fig~\ref{Figgrs} in order to determine the location of the outer horizon and the smallest root of~\eqref{as2} ($n=3$) we assumed the standard mass and the rotation parameter of the Kaluza-Klein BH to be $M/M_\odot =18.4$ and $a/r_g=0.98$, respectively, then we varied the mass within the limits~\eqref{pr3} to obtain plots of $\nu_U$ and $\nu_L$. This yields the numerical upper limit $c_{\text{max}}=2.06$. For the right panel of Fig~\ref{Figgro} we assumed the standard mass and the rotation parameter of the Kaluza-Klein BH to be $M/M_\odot =9.6$ and $a/r_g=1.00$, respectively. This yields the numerical lower limit $c_{\text{min}}=2.00$. Using~\eqref{ratio2} we obtain
\begin{equation}\label{ratio5}
0.00\lesssim \frac{Q^2}{4\pi\epsilon_0 G M^2}\lesssim 0.03.
\end{equation}
This interval is much smaller that the one we would have obtained had we treated this microquasar as a Kerr-Newman BH, that is, $0.00\lesssim Q^2/(4\pi\epsilon_0 G M^2)\lesssim 0.19$.

\section{Summary and conclusions \label{summary}}

We have obtained an analytic expression for the typical shadow radius with the inclination angle $i=\pi/2$. Using this equation, firstly we have shown that the typical shadow radius decreases with the increase of the angular momentum parameter and a fixed value of $b/c$. Secondly, we show that the typical shadow radius decreases by varying the parameters $c/b$ having fixed values for $a$. Thirdly, we have used the geometric-optics correspondence  and obtained a connection between the QNMs and the shadow radius.

The investigation of QPOs has led to restrict the values of the parameters, particularly of the ratio of the charge squared to the mass squared, of the Kaluza-Klein BH. A comparison with the Kerr-Newman BH reveals that the latter does not always convene to justify the occurrence of a lower QPO and of an upper QPO in a ratio of 3 to 2 observed in the three quasars considered in this work, while the Kaluza-Klein BH does.


\begin{acknowledgments}

M.G.-N. thanks School of Astronomy, Institute for Research in Fundamental Sciences (IPM), Tehran, Iran for their support. M.G.-N. and M.J. would like to thank Cosimo Bambi for helpful discussions during the early phase of this work.

\end{acknowledgments}


\begin{thebibliography}{99}

\bibitem{AbbottBH} B. P. Abbott et al. (LIGO Scientific and Virgo Collaborations), Phys. Rev. Lett. 116 (2016) 061102.
\bibitem{EHT1} The Event Horizon Telescope Collaboration, Astrophys. J. {\bf 875} (2019) L1.
\bibitem{EHT2} The Event Horizon Telescope Collaboration, Astrophys. J. {\bf 875} (2019) L2.
\bibitem{EHT3} The Event Horizon Telescope Collaboration, Astrophys. J. {\bf 875} (2019) L3.
\bibitem{EHT4} The Event Horizon Telescope Collaboration, Astrophys. J. {\bf 875} (2019) L4.
\bibitem{EHT5} The Event Horizon Telescope Collaboration, Astrophys. J. {\bf 875} (2019) L5.
\bibitem{EHT6} The Event Horizon Telescope Collaboration, Astrophys. J. {\bf 875} (2019) L6.  
\bibitem{Zhidenko1} A.~Zhidenko,  Class. Quant. Grav.  {\bf 21}, 273 (2004).
\bibitem{Blanchet} L. Blanchet, arXiv:1902.09801
\bibitem{Pretorius} F. Pretorius, Phys. Rev. Lett. {\bf95},121101 (2005).
\bibitem{Campanelli} M. Campanelli, C. O. Lousto, P. Marronetti and Y. Zwlochower, Phys. Rev. Lett. {\bf96}, 111101 (2006).
\bibitem{Baker} J. G. Baker et al., Phys. Rev. Lett. {\bf96}, 111102 (2006).
\bibitem{BertiCardosoWill} E. Berti, V. Cardoso and C. Will, Phys. Rev. D {\bf73}, 064030 (2006).
\bibitem{Regge} T. Regge and J. A. Wheeler, Phys. Rev. 108 (1957) 1063.
\bibitem{Zerilli} F. J. Zerilli, Phys. Rev. D {\bf2}, 2141 (1970).
\bibitem{1} E.~Berti and K.~D.~Kokkotas,  Phys.\ Rev.\ D {\bf 71}, 124008 (2005).
\bibitem{2} B. Mashhoon, Phys. Rev. D {\bf31}, 290 (1985).
\bibitem{3} R.~A.~Konoplya and A.~Zhidenko, Rev.\ Mod.\ Phys.\  {\bf 83}, 793 (2011).
\bibitem{4} V. Ferrari and B. Mashhoon, Phys. Rev. D {\bf30}, 295 (1984).
\bibitem{5} B. F. Schutz and C. M. Will, Astrophys. J. Lett. 291 (1985) L33.
\bibitem{6} S. Iyer and C. M. Will, Phys. Rev. D {\bf35}, 3621 (1987).
\bibitem{7} R. A. Konoplya, Phys. Rev. D 68 (2003) 024018.
\bibitem{8} S. Chandrasekhar and S. Detweiler, Proc. R. Soc. London, Ser. A {\bf344}, 441 (1975).
\bibitem{9} E. W. Leaver, Proc. R. Soc. London, Ser. A, {\bf402}, 285 (1985).
\bibitem{10} G. T. Horowitz and V. E. Hubeny, Phys. Rev D {\bf62}, 024027 (2000).
\bibitem{11} H. T. Cho et al., Adv. Math. Phys. 2012 (2012) 281705.
\bibitem{12} K. D. Kokkotas and B. G. Schmidt, Living Rev. Rel. 2 (1999) 2.
\bibitem{13} E. Berti, V. Cardoso and A. O. Starinets, Class. Quant. Grav. {\bf26}, 163001 (2009).
\bibitem{14} R. A. Konoplya and A. Zhidenko, Rev. Mod. Phys. {\bf83}, 793 (2011).
\bibitem{15} R.~A.~Konoplya, Z.~Stuchlík and A.~Zhidenko, Phys.\ Rev.\ D {\bf 98},  104033 (2018).
\bibitem{16} S. H. Hendi, A. Nemati, K. Lin, M. Jamil, Eur. Phys. J. C {\bf 80}, 296 (2020).
\bibitem{Synge66}  J.~L.~Synge,  Mon.\ Not.\ Roy.\ Astron.\ Soc.\  {\bf 131}, 463 (1966).
\bibitem{Luminet79}  J.-P.~Luminet, Astron.\ Astrophys.\  {\bf 75}, 228 (1979).
\bibitem{DeWitt73}  J. M. Bardeen, in Black Holes (Proceedings, Ecole d'Eté de Physique Théorique: Les Astres Occlus : Les Houches, France, August, 1972) edited by C.~DeWitt and B.~S.~DeWitt
\bibitem{r1} L. Amarilla, E. F. Eiroa, Phys. Rev. D {\bf85}, 064019 (2012).
\bibitem{r2}  A. Abdujabbarov, F. Atamurotov, Y. Kucukakca, B. Ahmedov, U. Camci, Astrophys. Space Sci. {\bf344}, 429 (2013).
\bibitem{r3} U. Papnoi, F. Atamurotov, S. G. Ghosh, B. Ahmedov, Phys. Rev. D {\bf90}, 024073 (2014).
\bibitem{r4} R. A. Konoplya, Phys. Lett. B {\bf804}, 135363 (2020).
\bibitem{r5} R. Shaikh, P. S. Joshi, JCAP {\bf10}, 064 (2019).
\bibitem{r6} D. Psaltis, Gen. Rel. Grav. {\bf51}, 137 (2019).
\bibitem{r7} S. Vagnozzi, L. Visinelli, Phys. Rev. D {\bf100}, 024020 (2019).
\bibitem{r8} R. Kumar, S. G. Ghosh, JCAP {\bf07} 053 (2020).
\bibitem{r9} R. Roy, S. Chakrabarti, Phys. Rev. D {\bf102}, 024059 (2020).
\bibitem{r10} V. Perlick, O. Y. Tsupko, G. S. Bisnovatyi-Kogan, Phys. Rev. D {\bf97},104062 (2018).

\bibitem{myshadow1} M.~Ghasemi-Nodehi, Z.~Li and C.~Bambi, Eur. Phys. J. {\bf C 75}, 315 (2015).
\bibitem{myshadow2} M.~Ghasemi-Nodehi, C.~Bambi,  Eur. Phys. J. {\bf C 76}, 290 (2016).
\bibitem{000}  K.~Hioki and K.~i.~Maeda, Phys.\ Rev.\ D {\bf 80} (2009) 024042
\bibitem{00} S.-W.~Wei, Y.-X.~Liu, R. B.~Mann, Phys. Rev. D {\bf 99}, 041303 (2019).
\bibitem{01}  S.-W.~Wei, Y.-C.~Zou, Y.-X.~Liu, R. B.~Mann, JCAP {\bf 1908}, 030 (2019).
\bibitem{111}  K.~Jusufi, M.~Jamil, P.~Salucci, T.~Zhu and S.~Haroon, Phys. Rev. D {\bf 100}, 044012 (2019).
\bibitem{22} T.~Zhu, Q.~Wu, M.~Jamil and K.~Jusufi, Phys.\ Rev.\ D {\bf 100}, 044055 (2019).
\bibitem{33} S.~Haroon, M.~Jamil, K.~Jusufi, K.~Lin and R.~B.~Mann, Phys.\ Rev.\ D {\bf 99}, 044015 (2019).
\bibitem{44} S.~Haroon, K.~Jusufi and M.~Jamil, Universe {\bf 6},  23 (2020).
\bibitem{55} K. Jusufi, M. Jamil, H. Chakrabarty, Q. Wu, C. Bambi, A. Wang, Phys. Rev. D \textbf{101}, 044035 (2020).
\bibitem{66}  C.~Bambi and K.~Freese, Phys.\ Rev.\ D {\bf 79}, 043002 (2009).
\bibitem{77}  C.~Bambi and N.~Yoshida,  Class.\ Quant.\ Grav.\  {\bf 27}, 205006 (2010).
\bibitem{88}  A.~Abdujabbarov, M.~Amir, B.~Ahmedov and S.~G.~Ghosh, Phys.\ Rev.\ D {\bf 93}, 104004 (2016).
\bibitem{99} M.~Amir and S.~G.~Ghosh, Phys.\ Rev.\ D {\bf 94}, 024054 (2016).
\bibitem{100} R.~Shaikh,  Phys.\ Rev.\ D {\bf 100}, 024028 (2019).
\bibitem{1111}  C.~Bambi, K.~Freese, S.~Vagnozzi and L.~Visinelli, Phys. Rev. D {\bf 100}, 044057 (2019).	
\bibitem{222} C.~Y.~Chen,  arXiv:2004.01440 [gr-qc].
\bibitem{333} R.~C.~Pantig and E.~T.~Rodulfo,  arXiv:2003.06829 [gr-qc].
\bibitem{444} L.~Amarilla and E.~F.~Eiroa,  Phys.\ Rev.\ D {\bf 85}, 064019 (2012).
\bibitem{Banerjee:2019nnj} I.~Banerjee, S.~Chakraborty and S.~SenGupta, Phys. Rev. D \textbf{101}, 041301 (2020).
\bibitem{Feng:2019zzn} X.~H.~Feng and H.~Lu,  arXiv:1911.12368 [gr-qc].
\bibitem{Zhang:2019glo}  M.~Zhang and M.~Guo,  arXiv:1909.07033 [gr-qc].
\bibitem{cardoso} V. Cardoso, A. S. Miranda, E. Berti, H. Witek, and V. T. Zanchin, Phys. Rev. D {\bf79}, 064016 (2009).
\bibitem{Hod:2017xkz} S.~Hod,  Phys.\ Lett.\ B {\bf 727}, 345 (2013).
\bibitem{Konoplya:2017wot} R.~A.~Konoplya and Z.~Stuchlík,Phys.\ Lett.\ B {\bf 771}, 597 (2017).
\bibitem{Wei:2019jve} S.~W.~Wei and Y.~X.~Liu, arXiv:1909.11911 [gr-qc].
\bibitem{Stefanov:2010xz} I.~Z.~Stefanov, S.~S.~Yazadjiev and G.~G.~Gyulchev, Phys.\ Rev.\ Lett.\  {\bf 104}, 251103 (2010).
\bibitem{Jusufi:2019ltj} K.~Jusufi, Phys. Rev. D \textbf{101}, 084055 (2020).
\bibitem{Liu:2020ola} C.~Liu, T.~Zhu, Q.~Wu, K.~Jusufi, M.~Jamil, M.~Azreg-Aïnou and A.~Wang, Phys. Rev. D {\bf 101}, 084001 (2020).
\bibitem{Jusufi:2020dhz} K.~Jusufi, Phys. Rev. D {\bf 101}, 124063 (2020).
\bibitem{geo} T.~Johannsen and D.~Psaltis, Astrophys. J. \textbf{718}, 446 (2010).
\bibitem{jet} R.~Craig Walker, P.~E.~Hardee, F.~B.~Davies, C.~Ly and W.~Junor, \emph{Astrophys. J.}  {\bf 855} (2018) 128.
\bibitem{Lund} E.~Lund, L.~Bugge, I.~Gavrilenko and A.~Strandlie, \emph{JINST} {\bf 4} (2009) P04001.
\bibitem{horne} J.H. Horne, G.T. Horowitz, Phys. Rev. D {\bf46}, 1340 (1992).
\bibitem{eiroa} L. Amarilla, E.F. Eiroa, Phys. Rev. D {\bf 87}, 044057 (2013).
\bibitem{twang} T. Wang, Nuc. Phys. B {\bf756}, 86 (2006).
\bibitem{long} F. Long, J. Wang, S. Chen, J. Jing, JHEP {\bf 10}, 269 (2019).
\bibitem{fin} F. Larsen, Nucl. Phys. B {\bf575}, 211 (2000).
\bibitem{xray} J. Zhu, A. B. Abdikamalov, D. Ayzenberg, M. Azreg-A\"{\i}nou, C. Bambi, M. Jamil, S. Nampalliwar, A. Tripathi, M. Zhou, Eur. Phys. J. C {\bf80}, 622 (2020).
\bibitem{KK} H.C. Lee (Ed.), \textit{An Introduction to Kaluza-Klein Theories}, (Singapore: World Scientific, 1984)
\bibitem{MMK} M. Azreg-Ainou, M. Jamil, K. Lin, Chin. Phys. C {\bf 44} 065101 (2020).
\bibitem{res}T. E.~Strohmayer, ApJ. Lett. \textbf{552}, L49 (2001).
\bibitem{qpos1}J. E.~McClintock et al., Class. Quantum Grav. \textbf{28}, 114009 (2011).
\bibitem{res2}R. Shafee, J. E. McClintock, R. Narayan, S. W. Davis, L.-X. Li, and R. A. Remillard, The Astrophysical Journal Letters \textbf{636}, L113 (2006).
\bibitem{res3}M. A. Abramowicz, V. Karas, W. Klu\'{z}niak, W. Lee and P. Rebusco, Publ. Astron. Soc. Japan, \textbf{55}, 467 (2003).
\bibitem{res4}J. Hor\'{a}k and V. Karas, A{\&}A, \textbf{451}, 377 (2006).
\bibitem{Kerr1}A. N.~Aliev and D. V.~Galtsov, Gen. Relativ. Gravit. \textbf{13}, 899 (1981).
\bibitem{qposknb}M.~Azreg-A\"{\i}nou, Int. J. Mod. Phys. D \textbf{28}, 1950013 (2019).
\bibitem{b1}L.D. Landau and E.M. Lifshitz, {Mechanics}, 3rd edition, (Pergamon Press, Oxford, 1976).
\bibitem{b2}A.H Nayfeh and D.T. Mook, {Nonlinear Oscillations}, (Wiley-VCH Verlag GmbH, New Jersey, 1995).
\bibitem{b3}A. Lindner and D. Strauch,
{A Complete Course on Theoretical Physics: From Classical Mechanics to Advanced Quantum Statistics},
(Springer Nature Switzerland AG, 2018).
\bibitem{b4}E.I. Butikov, Parametric resonance,
\href{\doibase 10.1109/5992.764219}{Computing in Science and Engineering (CiSE) May/June, 76 (1999)}.
\bibitem{fit}M. Kolo\v{s}, Z. Stuchl\'{i}k and A. Tursunov, Class. Quantum Grav. \textbf{32}, 165009 (2015).
\bibitem{fit2}M. Azreg-A\"{i}nou, Z. Chen, B. Deng, M. Jamil, T. Zhu, Q. Wu and Y.-K. Lim, Phys. Rev. D {\bf102}, 044028 (2020).


\end{thebibliography}
\end{document}